\documentclass[11pt, a4paper]{article}
\usepackage[margin=2cm]{geometry}
\usepackage{times,graphicx,amsmath,amssymb,pdflscape,natbib,authblk,lineno,setspace}
\usepackage[lofdepth,lotdepth]{subfig}
\usepackage{hyperref}

\newcommand{\Div}{{\boldsymbol{\nabla}}\cdot}
\newcommand{\Grad}{{\boldsymbol{\nabla}}}
\newcommand{\delsq}{\nabla^2}

\newcommand{\ldiff}[2]{\frac{\mathrm{D}{#1}}{\mathrm{D}{#2}}}

\newcommand{\strr}{\dot{\varepsilon}}
\newcommand{\Strr}{\dot{\boldsymbol{\varepsilon}}}
\newcommand{\infd}{\textrm{d}}
\newcommand{\xvec}{\boldsymbol{x}}

\newcommand{\vel}{\boldsymbol{v}}
\newcommand{\Pe}{\text{Pe}}
\newcommand{\waterunits}{OH$\,/10^6$Si}

\title{Grain-size dynamics beneath mid-ocean ridges: 
  Implications for permeability and melt extraction}
\author{Andrew J.~Turner$^1$, Richard F.~Katz$^1$ \& Mark D.~Behn$^2$}
\affil{$^1$ Department of Earth Sciences, University of Oxford, South Parks Road, Oxford OX1 3AN, United Kingdom}
\affil{$^2$ Department of Geology and Geophysics, Woods Hole Oceanographic Institution, 360 Woods Hole Road - MS \#22, Woods Hole, MA 02543, USA }
\affil{email: andrew.turner@earth.ox.ac.uk}

\begin{document}
\maketitle

\begin{abstract}
  Grain size is an important control on mantle viscosity and
  permeability, but is difficult or impossible to measure \textit{in
    situ}.  We construct a two-dimensional, single phase model for the
  steady-state mean grain size beneath a mid-ocean ridge.  The mantle
  rheology is modelled as a composite of diffusion creep, dislocation
  creep, dislocation accommodated grain boundary sliding, and a
  plastic stress limiter.  The mean grain size is calculated by the
  piezometric relationship of \cite{AustEv07}.  We investigate the
  sensitivity of our model to global variations in grain growth
  exponent, potential temperature, spreading-rate, and mantle
  hydration.  We interpret the mean mean grain-size field in the
  context of permeability.  The permeability structure due to mean
  grain size may be approximated as a high permeability region beneath
  a low permeability region. The transition between high and low
  permeability regions forms a boundary that is steeply sloped toward
  the ridge axis.  We hypothesise that such a permeability structure
  generated from the variability of the mean grain size may be able to
  focus melt towards the ridge axis, analogous to a
  \cite{sparks91}-type focusing.  This focusing may, in turn,
  constrain the region where significant melt fractions are observed
  by seismic or magnetotelluric surveys. This interpretation of melt
  focusing via the grain-size permeability structure is consistent
  with MT observation of the asthenosphere beneath the East Pacific
  Rise \citep{Baba06, Key13}.
\end{abstract}

\newpage

\section{Introduction}
\label{sec:intro}

Mid-ocean ridges (MOR) are a fundamental feature of terrestrial plate
tectonics and the simplest of the main tectono-volcanic systems.  The
asthenospheric dynamics beneath and near MORs are driven mostly by
spreading of lithospheric plates, which is a consequence of far-field
tectonic stresses (e.g.~slab pull). The passive asthenospheric flow
caused by imposed plate spreading is dominantly controlled by the
material properties of the asthenosphere and, in particular, its
viscosity.  Furthermore, asthenospheric flow beneath a ridge causes
melting; this melt segregates to fuel MOR volcanism and production of
oceanic crust.  Melt segregation is controlled by the permeability of
the partially molten asthenosphere.  Both mantle permeability and
viscosity are sensitive to mantle grain size, a key property that has
received little consideration in most previous models.

Grain size is a fundamental structural property of a polycrystalline
material that can vary in response to conditions including stress,
strain rate, temperature, and the presence of melt.  Grain size growth
and reduction are thought to be consequences of independent and
simultaneous processes \citep[e.g.][]{AustEv07, HallParmentier03}.  In
situations where these rates are balanced, a steady-state grain size
can be established.  However, predictions of grain dynamics are
complicated by the non-linear relationships between the grain size,
viscosity, and stress, which can lead to reinforcing feedbacks.

Ductile strain localisation is a well-studied example of a grain-size
feedback \citep{poirier80, jessell91, drury91, jin98, Braun99,
  montesi03, Berco03}. It occurs when the viscosity is positively
correlated with grain size. Deformational work reduces the local grain
size, which in turn reduces the viscosity. A decrease in viscosity
allows the local strain-rate to increase, which further reduces the
local grain size. This feedback mechanism is the basis for an
instability that can emerge from an inhomogeneous initial viscosity
and/or grain size field and lead to strain localisation.  In the
simple form discussed here, it does not rely on the presence of fluid
or melt.  However, strain localisation in the presence of melt may
lead to the generation of melt bands, which can lead to additional
feedbacks on the localization process \citep{katz06, rudge14}.

A second feedback in which grain size plays a role is associated with
reactive flow of magma through a permeabile mantle matrix
\citep{kelemen95, aharonov95}.  Magma rising under buoyancy is
undersaturated in SiO$_2$ and hence dissolves pyroxene and
precipitates olivine; this process leaves a dunite residue as evidence
of extensive reaction \citep{morgan03, morgan05}.  If pyroxene is a
pinning phase that limits the growth of olivine grains
\citep{Evans2001}, then reactive dissolution may enable more rapid
growth of olivine.  Since permeability depends on the square of grain
size \citep[e.g.][]{vonbargen86}, this would increase permeability,
strengthening channelisation and reactive dissolution, and enabling
further olivine grain growth \citep{braun04}.

These two examples of feedback mechanisms emphasise the importance of
grain-size variations in time and space for controlling the dynamics
of mantle processes.  Unfortunately, there are no direct measurements
of \textit{in situ} grain size in the Earth's mantle.  Mantle
xenoliths \citep{Twiss77, Lallemant80} and ophiolites \citep{braun04}
can provide estimates for the range of grain sizes in the upper
mantle, however, such studies provide no information about spatial
variations in grain size on the scale of mantle dynamics beneath a
ridge axis.  Moreover, it is difficult to assess how much these
samples have evolved during emplacment, and thus how representative
the recorded grain sizes are of normal mantle conditions.  Similarly
seismic attenuation, which is a strong function of grain size
\citep{karato03}, typically cannot resolve grain size variations on
the length-scales that are important for controlling ridge dynamics.

An alternative approach for assessing grain size variations in the
mantle is to couple numerical models with experimentally derived flow
laws and grain-size evolution models.  \cite{behn09} used this
approach to estimate grain size as a function of depth in the oceanic
upper mantle.  As part of their study they compared the models of
\cite{HallParmentier03} and \cite{AustEv07} with experimental data for
deformed wet and dry olivine. They found that the \cite{AustEv07}
model provided closer agreement with the laboratory
experiments. \cite{behn09} modeled grain size in a one-dimensional
vertical column with a composite rheology of dislocation and diffusion
creep. The steady-state grain size was calculated under the assumption
that a constant fraction of mechanical work acts to reduce grain
size. They found that the mean grain-size reaches a minimum of
15-20~mm at a depth of approximately 150~km.  They also found that the
structure of mean grain size is a good fit to the low seismic
shear-wave velocity zone in the upper oceanic mantle. They predicted
that dislocation creep is the dominant deformation mechanism for all
depths of the upper mantle.  However, \cite{behn09} did not calculate
the influence of mantle corner flow, and so the near-ridge strain-rate
structure was over-simplified.  Moreover, the assumption of a constant
fraction of mechanical work reducing the mean grain size, as opposed
to a fraction of dislocation work \citep{AustEv07}, removed a
potentially important coupling between the deformation mechanism and
mean grain size.

The goal of this study is to characterise the variations in grain-size
beneath a mid-ocean ridge, with particular focus upon the implications
for the permeability structure beneath the ridge. The permeability
structure is an important control on melt migration and has been
implicated as a key component in focusing of partial melt towards the
ridge axis. In such focusing models \citep[e.g.][]{sparks91,
  spiegelman93c, ghods00, hebert10}, the cold thermal boundary of the
lithosphere gives rise to a permeability barrier due to freezing of
melt within the pore space of the mantle. The buoyancy-driven vertical
transport of melt is inhibited beneath this barrier by a compaction
pressure gradient of that balances the melt buoyancy. If the thermal
boundary were perpendicular to the gravity vector, then melt would be
trapped at this boundary. However, the thermal boundary layer is
inclined towards the ridge axis, such that a component of the
compaction pressure gradient, which acts normal to the permeability
barrier, drives melt towards the ridge axis. However,
permeability-based models of melt focusing have yet to consider the
contribution of spatial variations in grain size beneath the ridge
axis. This leaves open the question of whether a gradient in grain
size can act as a permeability barrier and, if so, what effect would
this have on melt transport beneath a mid-ocean ridge.

In this study we construct a two-dimensional, single phase model for
the steady-state grain size beneath a mid-ocean ridge.  The model
employs a composite rheology of diffusion creep, dislocation creep,
dislocation accommodated grain boundary sliding, and a plastic stress
limiter. Our choice of rheology allows for a nonlinear coupling
between the mean grain size and strain rate; the mean grain size is
reduced by dislocation creep and grain boundary sliding, which then
affects the strain rate of diffusion creep and grain boundary sliding.
The mean grain size is calculated using the paleowattmeter model
\citep{AustEv07}. The dynamics of the model are described by Stokes
flow and the rheology is taken from experimental flow laws. 

The manuscript is organised as follows. We develop the model in
section~\ref{sec:model}. First the standard Stokes flow dynamics are
briefly outlined, then the composite rheology and mean grain-size
evolution model are presented in detail. Section~\ref{sec:model}
concludes by examining the sensitivity of the composite rheology to
variations in experimentally determined parameters for two different
grain boundary sliding parameterizations. In section~\ref{sec:results}
we present a reference case for grain-size dynamics beneath a
mid-ocean ridge and explore the sensitivity of the model to grain
boundary sliding parameters, water concentration, and parameter
perturbations within the mean grain-size evolution equation. In
section~\ref{sec:perm}, we investigate the influence of mean grain
size upon the permeability structure for an ultra-slow, slow, and fast
spreading-rate ridge. The permeability structure due to mean grain
size is then interpreted in the context of melt transport.

\section{Model}
\label{sec:model}

We consider a model of incompressible, constant density, Stokes flow
with variable viscosity.  The viscosity is associated with a set of
simultaneously active creep mechanisms with rates that depend on
pressure, temperature, strain rate and, distinct from most previous
work, the mean grain size. All of these fields are allowed to vary
spatially throughout the domain, however we examine only steady-state
solutions.

\subsection{Flow and thermal model}

In this context, conservation of mass, momentum, and energy reduce to
equations representing the incompressibility constraint, balance of
viscous stress with the pressure gradient, and balance of heat flow by
advection and diffusion. We neglect viscous dissipation of heat. The
governing equations, written in terms of nondimensional symbols, are
\begin{subequations}
  \label{eq:flow-temperature}
  \begin{align}
    \label{eq:gov_nondimensional1}
    \Div \vel &= 0,\\
    \label{eq:gov_nondimensional2}
    \Grad P - \Div 2\eta \Strr &= {\bf{0}},\\
    \label{eq:gov_nondimensional3}
    \Div \vel T - \Pe^{-1}\delsq T &= 0,
  \end{align}
\end{subequations}
where $\vel$ is the velocity, $P$ is the dynamic pressure, $\eta$ is
the effective viscosity, $T$ is potential temperature, and
$\Strr = \left( \Grad \vel + \left(\Grad \vel
  \right)^\intercal\right)/2$
is the strain rate tensor. The equations have been nondimensionalised
with the characteristic scales: plate speed $U_0$, domain height $H$,
viscosity $\eta_0$, pressure $P_0=\eta_0U_0/H$, and mantle potential
temperature $T_p$.  These scalings give rise to the dimensionless
Peclet number $\Pe=U_0H/\kappa$, a measure of the relative importance
of advective to diffusive heat transport ($\kappa$ is the thermal
diffusivity). The viscosity is capped at a dimensional value of
$10^{24}$~Pa-s to improve the efficiency of numerical solutions.

The domain is a rectangle with the left edge aligned vertically
beneath the spreading axis and the top boundary coincident with the
sea floor, perpendicular to the spreading axis.  The velocity and
temperature at the top boundary are set as
$\vel = [\tanh \left(2x/x_r \right),\,0]^\intercal$ and $T = 0$, where
$x_r$ is the width of distributed extension by normal faulting at the
ridge axis.  Here, $x_r$ is taken to be 4 km.  The temperature at the
bottom boundary is set to one to represent adiabatic inflow of ambient
mantle. The dynamic pressure is set to zero on the right boundary.
All other boundary conditions enforce zero gradient normal to the
relevant boundary.  The reflection boundary conditions on the vertical
boundary beneath the ridge axis represents an assumption of symmetry
across the ridge axis.  This assumption is invalid for ridges that
migrate in a reference frame fixed on the deep mantle, but here we
ignore complexities associated with ridge migration.

An explicit model for the viscosity $\eta$ is required to close the
system of equations \eqref{eq:flow-temperature}; this is developed in
the following section.

\subsection{Rheology}
\label{sec:rheology}

To derive an effective viscosity for
equation~\eqref{eq:gov_nondimensional2} we begin with the assumption
that various deformation mechanism are simultaneously active, and that
their individual strain rates sum to produce the total strain rate
\begin{equation}
  \label{eq:strr_sum}
  \Strr = \sum_k \Strr_k = \sum_k \frac{{\boldsymbol{\sigma}}}{2 \eta_k},
\end{equation}
where $k$ is an index corresponding to the deformation mechanism and
$\Strr_k$ is the strain rate tensor associated with the $k^\text{th}$
deformation mechanism.  Each mechanisms is driven by the total
deviatoric stress $\boldsymbol{\sigma}$ at a rate that is consistent
with its own viscosity $\eta_k$.  Equation~\eqref{eq:strr_sum} can be
rearranged to give the effective viscosity,
\begin{equation}
  \label{eq:visc_sum}
  \eta = \left(\sum_k \eta_k^{-1} \right)^{-1}.
\end{equation}
This harmonic sum represents the physical concept that deformation at
a point in the mantle is dominated by the mechanism with the lowest
viscosty, such that $\eta\approx\min_k\eta_k$.

Here we consider four mechanisms of rock deformation: dislocation
creep ($L$), diffusion creep ($D$), grain boundary sliding ($G$), and
a brittle plastic stress limiter ($B$) such that
$k = \left\lbrace L,D,G,B \right\rbrace$. The latter is described by a
Drucker-Prager yield criterion that may be written as
$\sigma_Y = C \cos \Phi + \bar{P} \sin \Phi$, where $\sigma_Y$ is the
scalar yield stress, $C$ is the cohesion, $\bar{P}$ is the total
pressure and $\Phi$ is the friction angle. The Drucker-Prager yield
criterion can be rewritten as a viscosity by using (\ref{eq:strr_sum})
to give \begin{equation}
  \label{eq:stress_limit}
  \eta_B = \frac{C \cos \Phi + \bar{P} \sin \Phi}{2 \strr}.
\end{equation}
The cohesion and friction angle are constants within the model
presented and the yield criterion is assumed to be independent of
grain size. The inclusion of a plastic deformation mechanism puts an
upper limit on the amount of stress a volume can support. An increase
in the percentage of plastic flow within a volume has the effect of
reducing the viscosity of the volume, such that $\eta = \eta_B$ when
$\sigma_{II} = \sigma_Y$, where $\sigma_{II}$ is the second invariant
of the deviatoric stress tensor ${\boldsymbol{\sigma}}$.

A general formulation of the viscosity for dislocation creep,
diffusion creep, and grain-boundary sliding includes an Arrhenius
factor, a power-law dependence on mean grain size $a$, a power-law
strain-rate dependence, and a power-law dependence on water
concentration $C_{OH}$,
\begin{equation}
  \label{eq:flow_law}
  \eta_k = A^{D,W}_{k} a^{{m_k}/{n_k}} \exp\left(\frac{E_k +
      \bar{P}V_k}{n_kRT}\right) \strr_{II}^{(1-{n_k})/{n_k}} C_{OH}^{-r_k/n_k} ,
\end{equation}
where $A^{D,W}_{k}$ is an experimentally determined prefactor for dry
or wet conditions, $E_k$ is the activation energy, $V_k$ is the
activation volume, $R$ is the universal gas constant, $\bar{P}$ is the
total pressure, $\strr_{II}$ is the second invariant of the composite
strain-rate, $\strr_{II} = \sqrt{\Strr:\Strr/2}$, and $C_{OH}$ is the
water content in units of \waterunits. The exponents $n_k$, $m_k$, and
$r_k$ control the sensitivity to strain rate, grain size and water
content, respectively.  $n_k \neq 1$ enforces a sensitivity to strain
rate, and is associated with non-Newtonian viscosity. A full list of
parameter values and units is provided in Table~\ref{tab:viscsymbols}.

\begin{table}[ht]
  \centering
  \begin{tabular}{llll}
    Symbol & units & Description & Value \\
    \hline
    $A_L^{D}$ & sec$^{-1}$ MPa$^{-n}$ & Dislocation prefactor &  $1.1 \times10^5$ \\
    $A_L^{W}$ & sec$^{-1}$ MPa$^{-n}$ & Dislocation prefactor for wet composition &  $1.6 \times10^4$ \\
    $E_L$ & J/mol & Dislocation activation energy & $5.3 \times10^5$ \\
    $V_L$ & m$^3/$mol & Dislocation activation volume & $1.6 \times10^{-5}$ \\
    $n_L$ & - & Dislocation Stress exponent & $3.5$ \\
    $m_L$ & - & Dislocation Grain-size exponent & $0$ \\
    $r_L$ & - & Dislocation water exponent &  $1.2$ \\
    $A_D^{D}$ & $\mu$m$^3$ sec$^{-1}$ MPa$^{-1}$ & Diffusion prefactor &  $1.59 \times10^9$ \\
    $A_D^{W}$ & $\mu$m$^3$ sec$^{-1}$ MPa$^{-1}$ & Diffusion prefactor for wet composition &  $2.5 \times10^7$ \\
    $E_D$ & J/mol & Diffusion activation energy & $3.75 \times10^5$ \\
    $V_D$ & m$^3$/mol & Diffusion activation volume & $4 \times10^{-6}$ \\
    $n_D$ & - & Diffusion Stress exponent & $1$ \\
    $m_D$ & - & Diffusion Grain-size exponent & $3$ \\
    $r_D$ & - & Diffusion water exponent &  $1$ \\
    $A^{D,W}_{G(H,HK)}$ & $\mu$m$^3$ sec$^{-1}$ MPa$^{-1}$ & GBS prefactor &  $10^{4.8},~6.5 \times 10^3$ \\
    $E_{G(H,HK)}$ & J/mol & GBS activation energy & $4.45 \times 10^5,~4 \times10^5$ \\
    $V_{G(H,HK)}$ & m$^3$/mol & GBS activation volume & $1.6 \times10^{-5},~1.6 \times10^{-5}$  \\ 
    $n_{G(H,HK)}$ & - & GBS Stress exponent & $2.9,~3.5$ \\
    $m_{G(H,HK)}$ & - & GBS Grain-size exponent & $0.7,~2$ \\
    $r_G$ & - & GBS water exponent &  $0$ \\       
    $C$ & Pa & Cohesion & $5 \times10^7$\\
    $\Phi$ & degrees & Friction angle & $30$ \\
    \hline
  \end{tabular}
  \caption{Symbols, units, and values for viscosity
    variables. Subscript $H$ and $HK$ denote the values stated by
    \cite{Hansen11} and \cite{hirth03} respectively.}
  \label{tab:viscsymbols}
\end{table}

Each of the three thermally activated deformation mechanisms has a
distinct combination of $m_k,n_k$. Diffusion creep is sensitive to
mean grain size ($m_D>0$) but insensitive to strain rate ($n_D=1$);
dislocation creep is independent of grain-size $(m_L=0)$ but is
non-Newtonian ($n_L>1$); dislocation-accommodated grain-boundary
sliding is both grain-size sensitive and non-Newtonian
($m_G>0,n_G>1$). The creep flow law parameters are taken from
laboratory experiments.  In particular, we adopt values from
\cite{hirth03} for diffusion and dislocation creep; for grain-boundary
sliding we consider parameter values from both \cite{hirth03} and
\cite{Hansen11}.  Values are given in Table~\ref{tab:viscsymbols}.

It is evident from Equation \ref{eq:flow_law} that to compute the
viscosity associated with diffusion creep and grain-boundary sliding
we require, as an input, the spatial distribution of mean grain size
$a(\boldsymbol{x})$.  In the next section we describe a theory for
dynamic grain size that completes our model.

\subsection{Mean grain size}

Following \cite{behn09}, we adopt the theory for grain size evolution
elaborated by \cite{AustEv07}. They assume that the rate of change of
mean grain size equals the difference between the rate of normal grain
growth and the rate of grain size reduction by recrystallization such
that at steady-state, the grain growth rate is equal to the rate of
grain size reduction.  Below, we consider normal grain growth and
dynamic recrystallisation in turn, and then discuss the combined
theory.

\subsubsection{Normal grain growth}

\cite{BurTur52} formulated a canonical model for normal grain-growth
kinetics. They hypothesise that grain--grain boundaries migrate,
changing grain sizes, due to the pressure difference between grains.
On the scale of individual, neighbouring grains, pressure differences
arise from differences in surface tension. Surface tension (and hence
pressure) is inversely proportional to the radius of curvature of the
grain boundary.  The pressure difference between two grains at their
boundary therefore causes atoms to preferentially migrate from smaller
to larger grains. Mean grain size increases while the surface area of
grain--grain boundaries decreases, leading to a reduction in the
stored energy of the system.

\cite{BurTur52} assumed that the free energy per unit area and the
mobility of grain--grain boundaries are independent of grain size.
The mean grain size therefore varies with time $t$ as
$a\propto t^{1/p}$, where $p$ is the grain growth exponent.
Differentiating and eliminating $t$ gives the rate of grain growth
$\infd a/\infd t \propto a^{1-p}/p$.  Since grain growth is also a
thermally activated process, its rate may be written as
\begin{equation}
  \label{eq:gs_growth}
  \dot{a}_\text{growth} = \frac{K_g a^{1-p}}{p}\exp
  \left(-\frac{E_g + \bar{P}V_g}{RT}\right).
\end{equation}
where $K_g$ is the constant of proportionality.

Following from the argument above, the theoretically determined value
for the grain growth exponent is $p = 2$ \citep[][and references
therein]{BurTur52, Hill65, Atkin88}. \cite{Atkin88} argued that $p=2$
is an idealised case for a single phase system. Experiments have found
that environmental factors including temperature, crystal composition,
and the presence of impurities, melt, and volatiles all may affect the
grain-growth exponent \citep[][and references therein]{Atkin88,
  Evans2001}. In the present manuscript, we employ an empirically
determined reference value $p=3$ \citep{Evans2001, behn09} and
consider the sensitivity of the results to different values of this
exponent.

\subsubsection{Grain-size reduction}

In the \cite{AustEv07} model of grain size reduction, the energy
required to create new grain boundaries by dynamic recrystalisation is
supplied by mechanical work. They postulate that some fraction of the
work is reversibly transferred into surface energy of new grain
boundaries, while the remainder is dissipated irreversibly as heat.
The volumetric mechanical work may be written as
$W = {\boldsymbol{\sigma}} : {\boldsymbol{\varepsilon}}$ and therefore
the volumetric work rate is
$\dot{W} = \dot{{\boldsymbol{\sigma}}}: \boldsymbol{\varepsilon} +
\boldsymbol{\sigma}:\Strr$.
In the steady-state limit, $\dot{{\boldsymbol{\sigma}}} = 0$.  Thus,
the work rate per unit volume reduces to
$\dot{W} = {\boldsymbol{\sigma}}:\Strr$.

The change in internal energy per unit volume $\dot{\mathcal{E}}$ may
be written as,
\begin{equation}
  \frac{\partial \mathcal{E}}{\partial t}
  \propto \frac{\gamma}{V}\frac{\partial S}{\partial t} 
  =\frac{\gamma}{V}\frac{\partial S}{\partial a}
  \frac{\partial a}{\partial t},
\end{equation}
where $\gamma$ is the surface energy per unit area of the grain-grain
boundaries, and $S$ and $V$ are the surface area and volume of a grain
with characteristic size $a$. In this case, we can write
$\partial S/\partial a \propto a$ and $V \propto a^3$, thus the change
in energy per unit volume associated with grain size reduction is
$\dot{\mathcal{E}} = -c\gamma\dot{a}/a^2$, where $c$ is a
dimensionless constant associated with the ratio of surface area to
volume for a typical grain. For spherical and cubic grains $c = 6$ or
$c=12$, respectively.

Of the total rate of mechanical work,
$\dot{W}={\boldsymbol{\sigma}}:\Strr$, some fraction $\beta$ is
accomplished by processes that depend on the movement of dislocations
through the crystalline lattice (dislocation creep ($L$) and
dislocation-accommodated grain boundary sliding ($G$)).  In
particular,
\begin{equation}
  \label{eq:beta_def}
  \beta = \frac{\dot{W}_L + \dot{W}_G}{\dot{W}} = 
  \frac{\eta}{\eta_L} + \frac{\eta}{\eta_G},
\end{equation}
where we have used equations~\eqref{eq:strr_sum} and
\eqref{eq:visc_sum} to write the work rate in terms of viscosities. Of
the portion $\beta$ of the total work rate, some fraction $\lambda$
goes into the grain size reduction and the rest, $1-\lambda$, is
dissipated as an irreversible increase in entropy.

Using this approach \cite{AustEv07} arrived at the rate equation for
reduction of mean grain size due to the rate of mechanical work as
\begin{equation}
  \label{eq:mech_red_term}
  \dot{a}_\text{reduction} = - \left(\lambda\beta\boldsymbol{\sigma}
    :\Strr\right)\left(\frac{a^2}{c\gamma}\right).
\end{equation}
The quantity in the first set of parentheses represents the rate of
work per unit volume acting to reduce mean grain size; the quantity in
the second set of parentheses represents the energetic cost per unit
volume of grain size reduction.

\subsubsection{Evolution of mean grain size}

Combining the rates of grain size growth \eqref{eq:gs_growth} and
reduction \eqref{eq:mech_red_term} additively \citep{AustEv07} along
trajectories of mantle flow gives the full equation for grain size
evolution as
\begin{equation}
  \label{eq:mean_grain_di}
  \ldiff{a}{t} = 
  - \frac{\beta\lambda\boldsymbol{\sigma}:\Strr}{c\gamma} a^2
  + \frac{K_g}{p}\exp\left(-\frac{E_g + \bar{P}V_g}{RT}\right)a^{1-p}.
\end{equation}

Equation~\eqref{eq:mean_grain_di} is put into non-dimensional form
using the following characteristic scales: $\vel = U_0\vel'$,
$\xvec = H\xvec'$, $\eta = \eta_0\eta'$,
$\boldsymbol{\sigma} = \eta_0U_0/H\boldsymbol{\sigma}'$, and
$a = a_0 a'$. The mean grain size may vary over many orders of
magnitude and therefore we make the substitution
$\mathcal{A} = \ln a'$ to improve numerical stability.  With these
modifications and assuming that grain size is in a quasi-steady state,
equation~(\ref{eq:mean_grain_di}) becomes
\begin{equation}
  \label{eq:grain_rate_nondi}
  \vel \cdot \Grad \mathcal{A}  +
  \mathcal{D}\strr_{II}^2\eta\beta\exp\left(\mathcal{A}\right) - 
  \mathcal{G}\exp\left(-\frac{E_g + \bar{P}V_g}{RT}
    - \mathcal{A}p  \right) = 0,
\end{equation}
where
\begin{equation}
  \label{eq:grain_coefs}
  \mathcal{D} = \frac{4\lambda\eta_0U_0a_0}{c\gamma H},\;\;\;
  \mathcal{G} = \frac{K_gH}{pa_0^pU_0},
\end{equation}
are the nondimensional coefficient of grain reduction and grain
growth, respectively. All symbols in
equation~\eqref{eq:grain_rate_nondi} are dimensionless except those in
the Arrhenius exponent. Parameter values and units associated with the
model for grain size evolution are given in
table~\ref{tab:grainsymbols}.

\begin{table}[ht]
  \centering
  \begin{tabular}{llll}
    Symbol & units & Description & Reference value \\
    \hline
    $\lambda$ & - & Fraction of dislocation work to grain size reduction & $1$  \\
    $c$ & - & Geometric factor & $12$\\
    $\gamma$ & J m$^{-2}$  & Surface energy at grain--grain contacts & $1$ \\
    $K_g$ & m$^{p}$ s$^{-1}$ & Grain-growth prefactor & $10^{-5}$ \\
    $p$ & - & Grain-growth exponent & 3 \\
    $E_g$ & J/mol & Grain-growth activation energy & $3.5\times10^5$ \\
    $V_g$ & m$^3$/mol & Activation volume for grain growth & $8 \times 10^{-6}$ \\
    \hline
  \end{tabular}
  \caption{Symbols, units and test values for grain evolution equation.}
  \label{tab:grainsymbols}
\end{table}

The grain growth and reduction parameters are dimensionless numbers
formed by taking the ratio of grain-growth and grain-reduction
prefactors to the characteristic scale associated with grain size
advection. Their size is not particularly meaningful, however, because
viscosity, strain rate, temperature, and grain size vary drastically
throughout the domain. Rates for grain growth and reduction are
similarly variable and hence characteristic scales are not available.
It is notable, however, that $U_0$ appears in the numerator of
$\mathcal{D}$ and in the denominator of $\mathcal{G}$. This suggests
that with increasing spreading rates, grain size reduction should
dominate over grain growth; we expect steady-state, mean grain size to
decrease with increasing spreading rate.

\subsection{Parameter values and deformation mechanisms}
\label{sec:param_values}

The governing partial differential equations
\eqref{eq:flow-temperature} and \eqref{eq:grain_rate_nondi} form a
closed system with the rheological model of
section~\ref{sec:rheology}. However, there is considerable uncertainty
in the values of laboratory-derived parameters associated with
dislocation accomodated grain-boundary sliding. In this section, we
use deformation-mechanism maps to highlight the physical consequences
that arise from the uncertainties in these parameters.

\cite{Hansen11} and \cite{hirth03} both estimated GBS parameter values
from laboratory data and obtained different results. In particular,
their values for the grain size exponent $m_G$ and the stress exponent
$n_G$ are in disagreement. \cite{Hansen11} obtained
$m_{G(H)} = 0.7 \pm 0.1$ and $n_{G(H)} = 2.9 \pm 0.3$. \cite{hirth03}
estimated $n_{G(H)} = 3.5 \pm 0.3$ and asserted that $m_{G}$ is
between 1 and 2; we therefore consider three values
$m_{G(HK)} = \{1, 1.5, 2\}$ and test the sensitivity of the dominant
deformation mechanism to the grain size exponent in the GBS
constitutive equation~\eqref{eq:flow_law}.  Note that we use eqn.~A4
in the appendix of \cite{Becker06} to account for the co-variation of
the viscous prefactor with stress and/or grain exponent.

\begin{figure*}
  \center
  \includegraphics[width = 12cm]{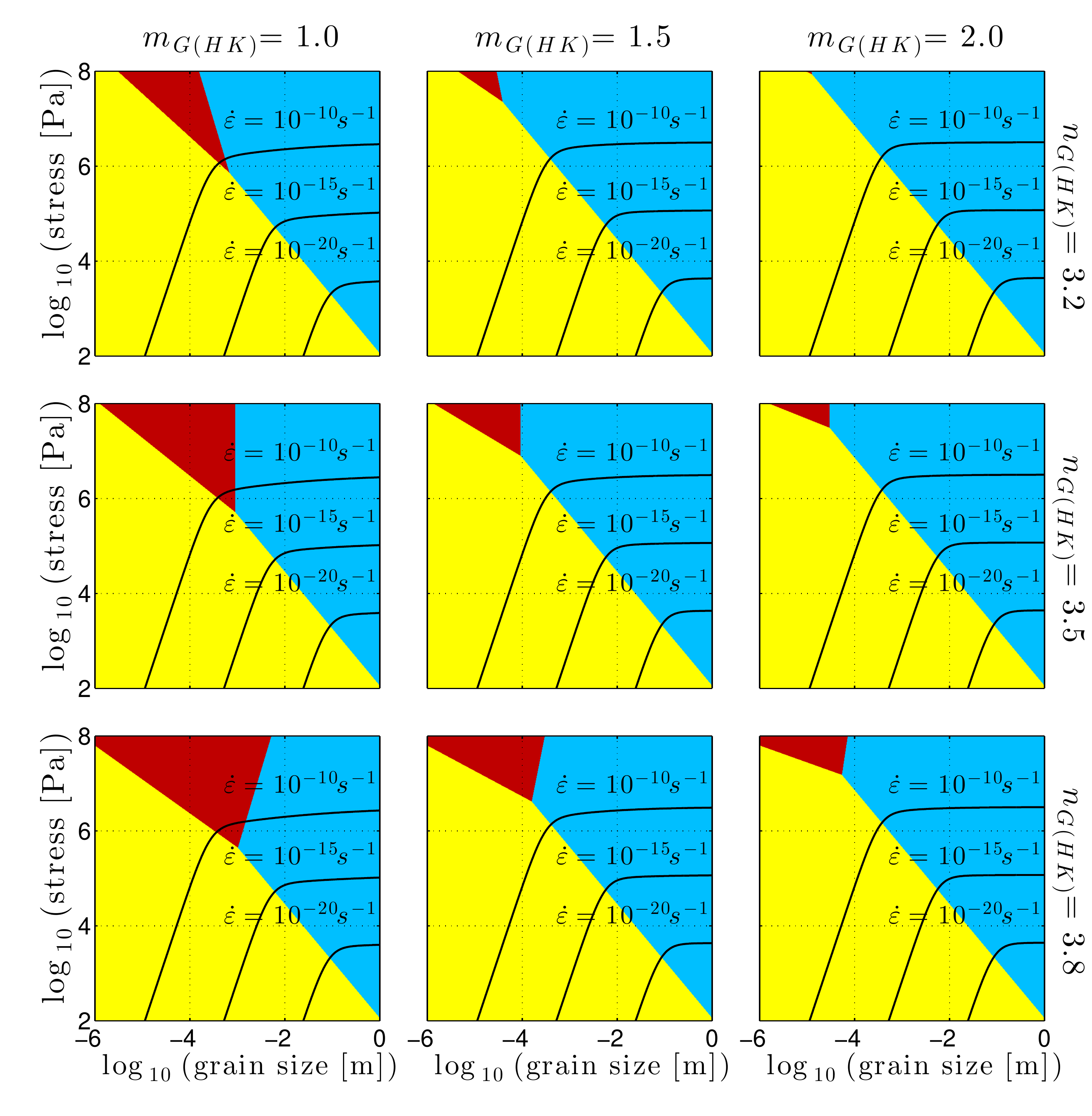}
  \caption{Deformation mechanism maps at a pressure and temperature of
    1~GPa and 1350$^\circ$C for the parameters found in
    \cite{hirth03}. Diffusion creep, grain boundary sliding, and
    dislocation creep are shown in yellow, red, and blue respectively.
    Contours for constant strain-rate have been added in black for
    rates of $10^{[-10,-15,-20]}$ s$^{-1}$. }
\label{fig:def_map_series_HK}
\end{figure*}

\begin{figure*}
  \center
  \includegraphics[width = 12cm]{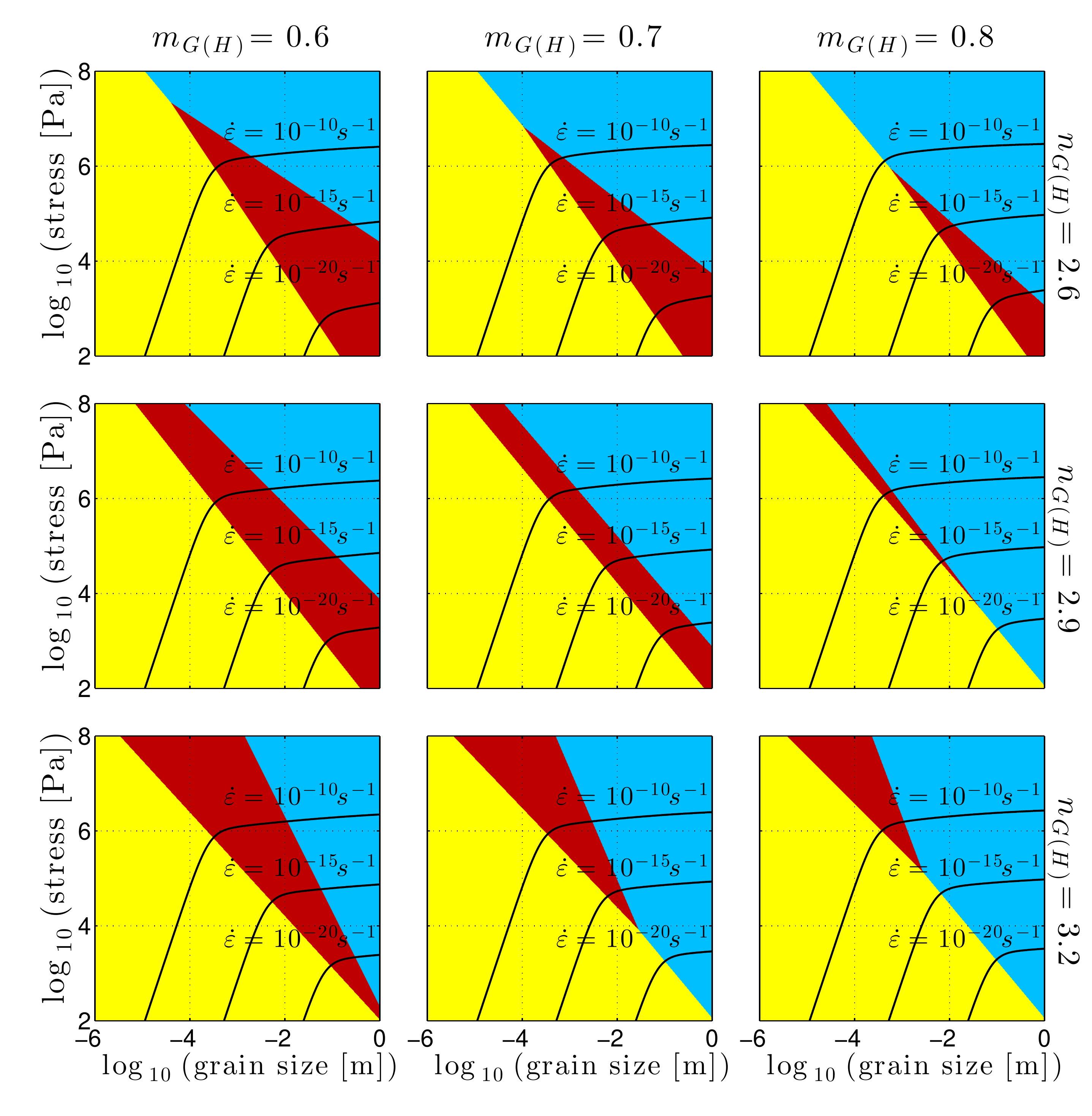}
  \caption{Deformation mechanism maps at a pressure and temperature of
    1~GPa and 1350$^\circ$C for the parameters found in
    \cite{Hansen11}. Diffusion creep, grain boundary sliding, and
    dislocation creep are shown in yellow, red, and blue respectively.
    Contours for constant strain-rate have been added in black for
    rates of $10^{[-10,-15,-20]}$ s$^{-1}$.}
  \label{fig:def_map_series_H}
\end{figure*}

Figures \ref{fig:def_map_series_HK} and \ref{fig:def_map_series_H}
show maps of the dominant deformation mechanism (that with the largest
contribution to the overall strain rate), as a function of stress and
grain size, for parameter values obtained from \cite{hirth03} and
\cite{Hansen11}, respectively.  Deformation mechanisms are computed at
1~GPa (about 30~km depth) and 1350~$^\circ$C. Solid black lines are
contours of constant strain rate appropriate for asthenospheric flow
beneath mid-ocean ridges.  Each plot is associated with a unique set
$(m_G,n_G)$, as labelled on the column and row of the plot.  All other
empirical values used in the flow law \eqref{eq:flow_law} are held
constant, including all experimentally determined values for diffusion
creep and dislocation creep. Regions dominated by diffusion creep,
dislocation creep, and grain boundary sliding are shown in yellow,
blue, and red, respectively.

For strain rates relevant for asthenospheric flow, the calculations
based on \cite{hirth03} predicted little contribution from
GBS. Therefore, if the Hirth \& Kohlstedt parameters are appropriate,
we do not expect sub-ridge grain size and flow to be sensitive to the
value of the GBS parameters used.  However, based on the parameters
from \cite{Hansen11}, GBS will be dominant over some range of
stress--grain size conditions relevant to mid-ocean ridges. The
details of its contribution will depend on the values of $m_{G(H)}$
and $n_{G(H)}$ that are chosen.

In constructing a reference model in the next section, we adopt the
parameter values suggested by \cite{Hansen11}.  The \cite{Hansen11}
parameters have the advantage that the values have been calibrated to
a single experimental set-up, the activation energy was measured, and
modern analytical techniques were used to measure grain size.

\section{Results}
\label{sec:results}

\subsection{Reference case}

We use the model described above to investigate the grain-size
dynamics beneath a mid-ocean ridge. We begin by considering how the
mean grain-size field, and dependent variables, behave for the
reference parameters shown in tables \ref{tab:viscsymbols} \&
\ref{tab:grainsymbols}. For the reference case we use the grain
boundary sliding values of \cite{Hansen11} and a half spreading-rate
of $U_0 = 2$~cm/a.  A half rate of 2~cm/a is typical of slow-spreading
ridges such as the mid-Atlantic ridge.

\begin{figure*}
  \center
  \includegraphics[width= \textwidth]{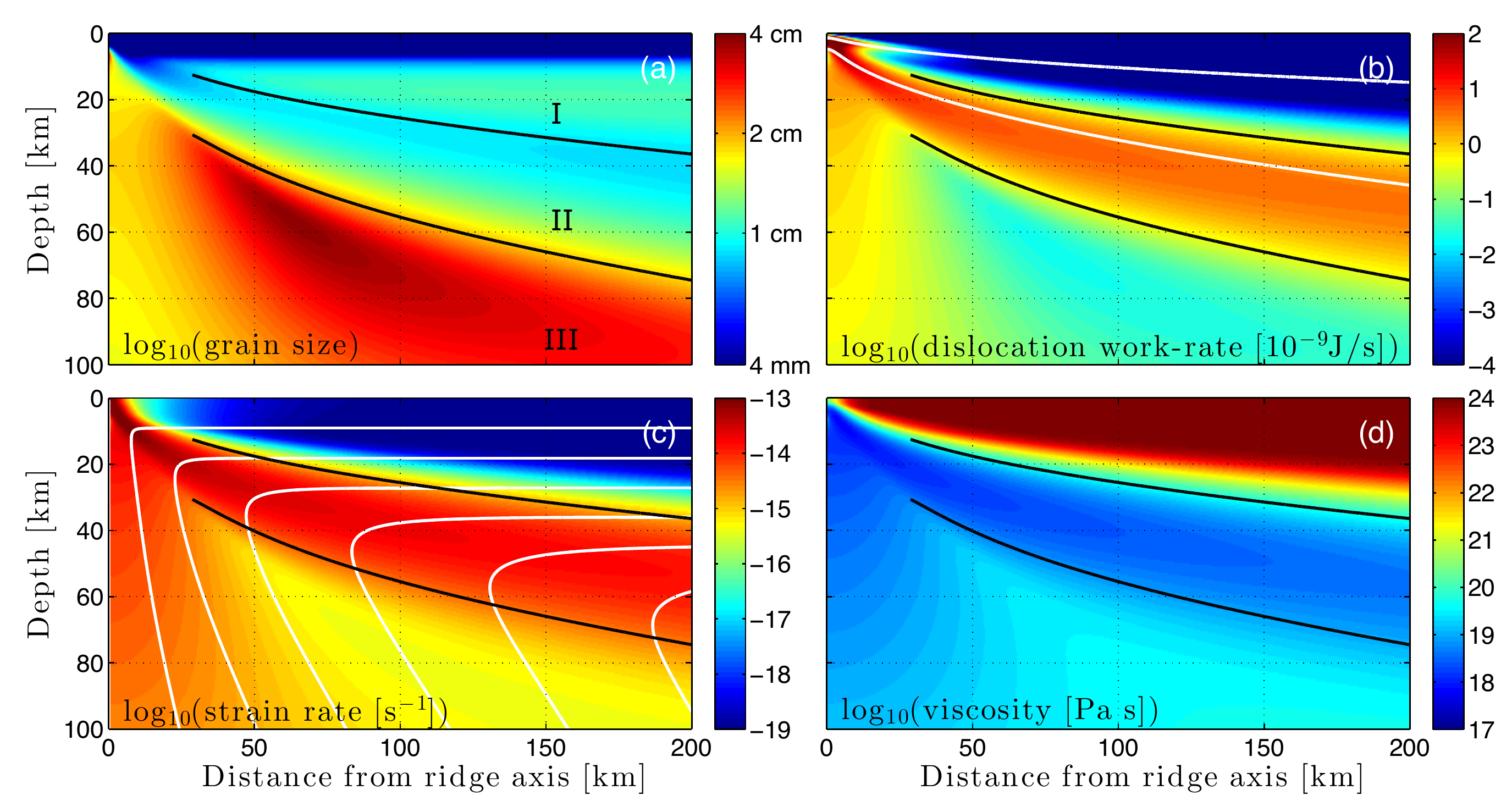}
  \caption{Representative fields for a half spreading rate of $U_0 =
    2$~cm/a. Panel (a): Grain size field. The colour-bar is truncated
    to show grain sizes from 4~mm to 4~cm. 
    Panel (b): Dislocation work-rate used to create new grain
    boundaries. White lines show contours for 1250$^\circ$C and
    600$^\circ$C. Panel (c): Second invariant of strain-rate. White
    contours show streamlines. Panel (d): Viscosity. 
    For all panels the black lines show the
    interface between regions I, II and III (see text for details)}
  \label{fig:rep_fields}
\end{figure*}

The mean grain size and associated fields are shown for the
reference parameters in Figure~\ref{fig:rep_fields}. Panel (a)
shows the mean grain size. The colour-bar has been truncated to
highlight grain sizes of 4~mm to 4~cm. Panel (b) shows the rate of
work per unit volume acting to reduce the mean grain size
($4\eta\strr^2\beta\lambda$); we call this the dislocation work-rate,
as it denotes the fraction of work done by processes that depend on
the movement of dislocations through the crystalline lattice.  White
lines display contours of potential temperature at 1250$^\circ$C and
600$^\circ$C. Panel (c) shows the second invariant of the strain-rate
tensor. White lines are streamlines of the solid flow. Panel (d) shows
the composite viscosity. For the purpose of explanation, the grain
size field may be roughly separated into three regions (I, II, III),
as shown in panel (a) of Figure~\ref{fig:rep_fields}. We now consider
each region in turn.

Region I represents the conductively cooled lithosphere, which
contains temperatures $<600^\circ$C. At such low temperatures
($T \lesssim 0.5T_m$, where $T_m$ is the melting temperature) viscous
creep mechanisms are not active, and the grain-size evolution model
used here does not apply.  In region I, cold temperatures drive both
the grain growth and reduction rates to zero. Therefore advection of
grain size becomes the dominant process. For negligible growth and
reduction rates we expect $\vel\cdot\Grad\mathcal{A}\approx0$ and
hence that grain size is approximately constant along
streamlines. Streamlines in the lithosphere are horizontal, leading to
constant grain size with distance from the ridge axis at depths where
the thermal profile results in negligible growth and reduction
rates. The typical grain size of region I is 1 cm with a minimum grain
size of the order of 10~$\mu$m near the surface.

Region II is an area of active deformation and contains the greatest
average strain-rates.  As such, the mean grain-size reduction term in
equation~(\ref{eq:grain_rate_nondi}) takes a maximum value in this
region. The grain growth-rate varies across region II due to the
temperature increase with depth. Generally, the temperature in region
II is greater than 1250$^\circ$C, which results in a large grain
growth rate. However, near the transition from region II to I, the
temperature drops sharply while the dislocation work-rate remains
high. This can be seen in panel (b) of Figure~\ref{fig:rep_fields},
where the colour indicates the magnitude of dislocation work-rate and
the white contours display the 1250$^\circ$C and 600$^\circ$C
isotherm. A combination of the dislocation work-rate structure and
thermal structure leads to a vertical gradient to the mean grain-size
within region II. The mean grain-size varies from approximately 2~cm
at the transition between regions II \& III to 6~mm at the transition
between regions I \& II.  Note that the small grain sizes near the
surface in region I are actually generated in region II.  As
immediately beneath the ridge axis the mantle flow must turn a tight
corner, and so here the dislocation work-rate is high
(Figure~\ref{fig:rep_fields}b).

Region III undergoes relatively slow deformation at high temperature.
As a result, diffusion creep becomes increasingly significant with
depth as both $\beta$ and $\strr_{II}$ decrease in the grain-size
reduction term.  The grain growth term remains uniformly high, due to
a high potential temperature in region III.  The combination of these
effects allows the mean grain size to grow most rapidly in region III
and achieve the largest mean grain size over the whole domain. The
typical mean grain size for region III is of the order of 2~cm,
ranging from approximately 1.5~cm to 4~cm.

As highlighted in Section~\ref{sec:param_values}, the parameters for
grain boundary sliding are uncertain and we therefore consider the
sensitivity of the reference model with respect to the grain boundary
sliding parameters. Near the ridge axis, the consequences of such
changes are subtle. Therefore, we use an extended domain to
demonstrate the implications of different choices for the grain
boundary sliding parameters at greater distances from the axis.

\subsection{Grain boundary sliding parameters}
\label{sec:sens_GBS_para}

Grain boundary sliding is the only deformation mechanism that gives
rise to a direct coupling between mean grain size and strain rate
within our model; it is therefore important to determine whether the
mean grain-size field is sensitive to the choice of grain boundary
sliding parameters. Here we investigate the dominant deformation
mechanism as a function of space beneath a mid-ocean ridge for the
grain boundary sliding parameters found by \cite{Hansen11} and
\cite{hirth03}. In addition, we evaluate and compare the strain rate
of each deformation mechanism within our composite rheology. We also
investigate the influence of grain boundary sliding parameters upon
the predicted mean grain-size field.

\begin{figure*}
  \center
  \includegraphics[width = 15.5cm]{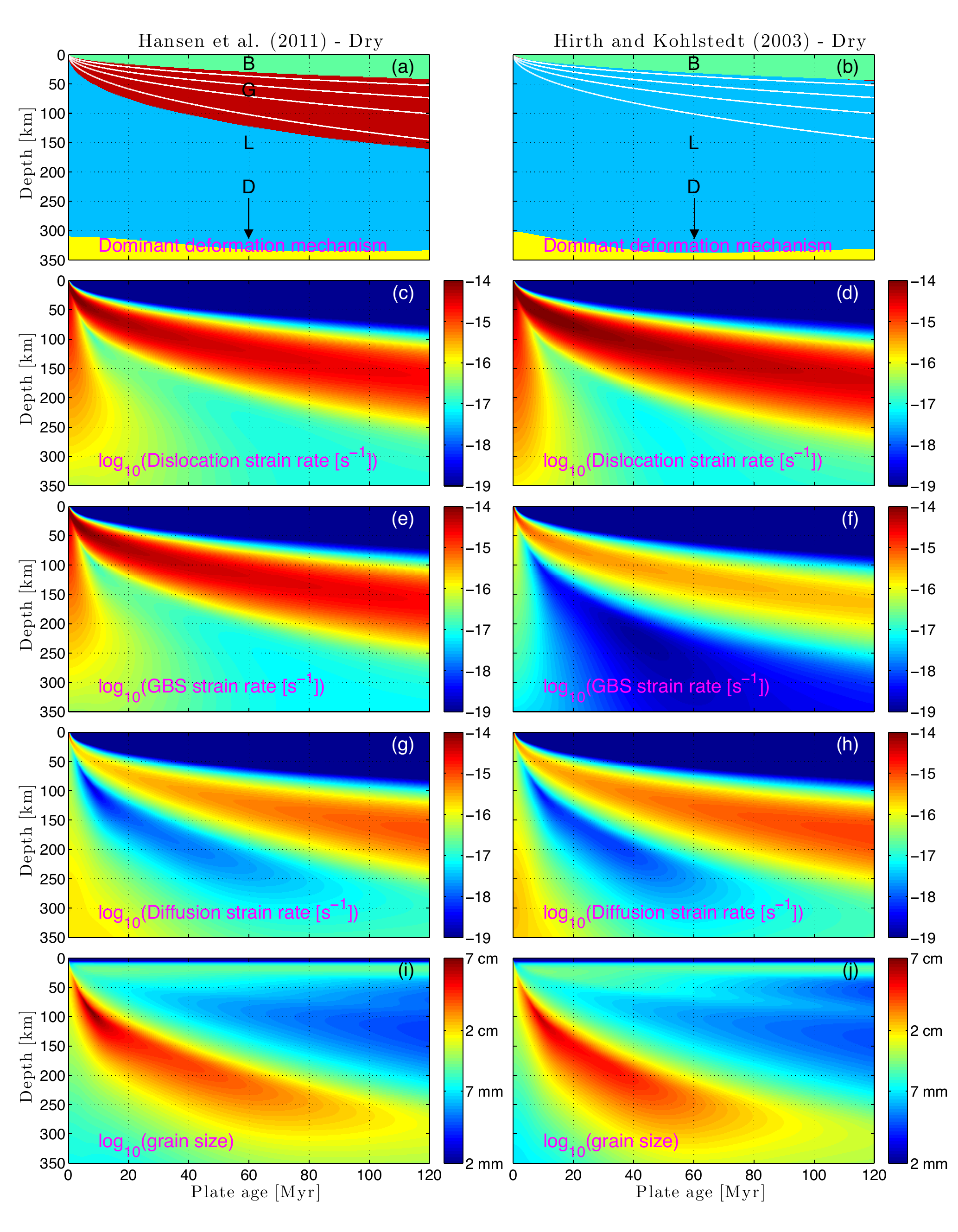}
  \caption{All panels are for a spreading-rate of $U_0 = 2$~cm/a.
    The left or right column use the \cite{Hansen11} or \cite{hirth03}
    parameters for grain boundary sliding.  Panels (a) \& (b): The
    dominant deformation mechanism. Plasticity (B), diffusion creep
    (D), grain boundary sliding (G), and dislocation creep (L) are
    shown in green, yellow, red, and blue respectively. Isotherms are
    shown for $T = \left\lbrace 1200, 1000, 800, 600
    \right\rbrace^\circ$C in white. Panels (c) \& (d): $\log_{10}($
    Dislocation strain rate $[s^{-1}])$. Panels (e) \& (f):
    $\log_{10}($ GBS strain rate $[s^{-1}])$ . Panels (g) \& (h):
    $\log_{10}($ Diffusion strain rate $[s^{-1}])$. Panels (i) \& (j):
    $\log_{10}($ grain size $)$. }
  \label{fig:domin_HKH_map}
\end{figure*}

We consider a large domain of 350~km depth and 2400~km width from the
ridge axis (corresponding, at $U_0=2$~cm/a to a maximum plate age of
120~Myr) in Figure \ref{fig:domin_HKH_map}. The two columns differ
only by the grain boundary sliding parameters used: the left column
uses the parameters determined by \cite{Hansen11} and the right column
uses the parameters from \cite{hirth03}. Panels (a) \& (b) display the
dominant deformation mechanism. Regions where plasticity (B),
diffusion creep (D), dislocation creep (L) and grain boundary sliding
(G) dominate are shown in green, yellow, blue and red respectively.
Isotherms are shown for $T_p = \{1200, 1000, 800, 600\}^\circ$C in
white. Panels (c)-(h) allow comparison between the magnitude of strain
rate in each creep component; dislocation creep is shown in panels (c)
\& (d), grain boundary sliding is shown in panels (e) \& (f), and
diffusion creep is shown in panels (g) \& (h). Panels (i) \& (j) show
the resulting mean grain size field.

Figure \ref{fig:domin_HKH_map} shows that the regions characterised by
plastic deformation and diffusion creep are unaffected by the grain
boundary sliding parameters. Diffusion creep controls the rheology at
depths greater than $\sim$300~km; plastic deformation dominates at
temperatures less than 600$^\circ$C. At depths below the 600$^\circ$C
isotherm and above 300~km depth, the dislocation-dependent deformation
mechanisms determine the rheology. As expected based on
section~\ref{sec:param_values}, the \cite{Hansen11} parameters enhance
grain boundary sliding, giving it a large region of dominance from the
middle to the base of the lithosphere. This region is located between
the 600$^\circ$C and 1200$^\circ$C isotherms. No such region is
predicted from the \cite{hirth03} parameters.

For the \cite{Hansen11} parameters, the dislocation-related strain
rate is split approximately evenly between dislocation creep and grain
boundary sliding (Figure~\ref{fig:domin_HKH_map} panel (c) and (e)).
In contrast, for the parameters of \cite{hirth03}, the dislocation
creep rate is 1--2 orders of magnitude greater than that of grain
boundary sliding. Moreover, under the \cite{hirth03} parameters, our
model predicts a greater strain rate for diffusion creep than grain
boundary sliding in the the majority of the asthenospheric mantle
beneath a mid-ocean ridge (Figure \ref{fig:domin_HKH_map} panel (f)
and (h)).

Despite these differences, the mean grain-size structure is relatively
insensitive to the grain boundary sliding parameters in the range
considered here. The maximum difference in mean grain size is less
than 80\% between models using the two parametrisations of grain
boundary sliding. For comparison, the spatial variation of mean grain
size within the domain for a single parametrisation is greater than
three orders of magnitude. All of these results, however, are obtained
under the assumption that the mantle is anhydrous. The presence of
water in olivine is known to weaken creep deformation.  In the next
section we consider how inclusion of a mantle water content that
depends only on depth influences the composite rheology and the flow.

\subsection{The effects of water}
\label{sec:water}

An important effect of water is to lower the solidus temperature of
the mantle, which allows partial melting to occur at a greater depth
than for anhydrous mantle. During mantle melting, water behaves as an
incompatible element \citep{hirth96}; the deepest, incipient melts are
highly enriched in water. At depths where the upwelling mantle is
above the anhydrous solidus, the solid residuum of melting is almost
completely dehydrated. Given these considerations, the concentration
of water in the mantle may be a simple function of depth, near the
ridge axis \citep[e.g.][]{braun00}. We parametrise the concentration
of water in the mantle as 
\begin{equation}
  \label{eq:waterprofile}
  C_{OH}(z) = \begin{cases}
    0 & \text{if $z < z_d$} \\
    C_{OH}^{\text{max}} (z-z_d) / (z_w-z_d) & \text{if $z_d < z < z_w$} \\
    C_{OH}^{\text{max}} & \text{if $z > z_w$} 
  \end{cases}
\end{equation}
where $C_{\text{OH}}^{\text{max}}$ is the maximum water concentration,
$z_w$ is the depth at which the mantle crosses the wet solidus, and
$w_d$ is the depth of the dry solidus.

The water concentration of the mantle beneath a mid-ocean ridge is
constrained to be below 2000 \waterunits{}, with current estimates
being $810 \pm 490$ \waterunits{} \citep{hirth96}. In this paper we
will consider values of $C_{\text{OH}}^{\text{max}} = \{ 750, 1500 \}$
\waterunits{} . These values approximately correspond to the mean and
upper limit of water concentration estimates. Note that the dominant
deformation mechanism of the upper mantle beneath a mid-ocean ridge is
dislocation creep with a viscosity proportional to $C_{OH}^{1.2/3.5}$
(see equation \ref{eq:flow_law}).  Therefore, differences between
$C_{\text{OH}}^{\text{max}} = 1300$ and 1500 or
$C_{\text{OH}}^{\text{max}} = 750$ and 810 \waterunits{} are
negligible.

As discussed above, among the deformation mechanisms considered here,
the parameters for grain boundary sliding are the least well
constrained. Currently there is no evidence that grain boundary
sliding is affected by the presence of water \citep{hirth03}.
Therefore, we set $A_G^W = A_G^D$ and $r_G = 0$.  Consequently, grain
boundary sliding becomes sub-dominant in regions where water is
present.

\begin{figure*}
  \center
  \includegraphics[width = 15.5cm]{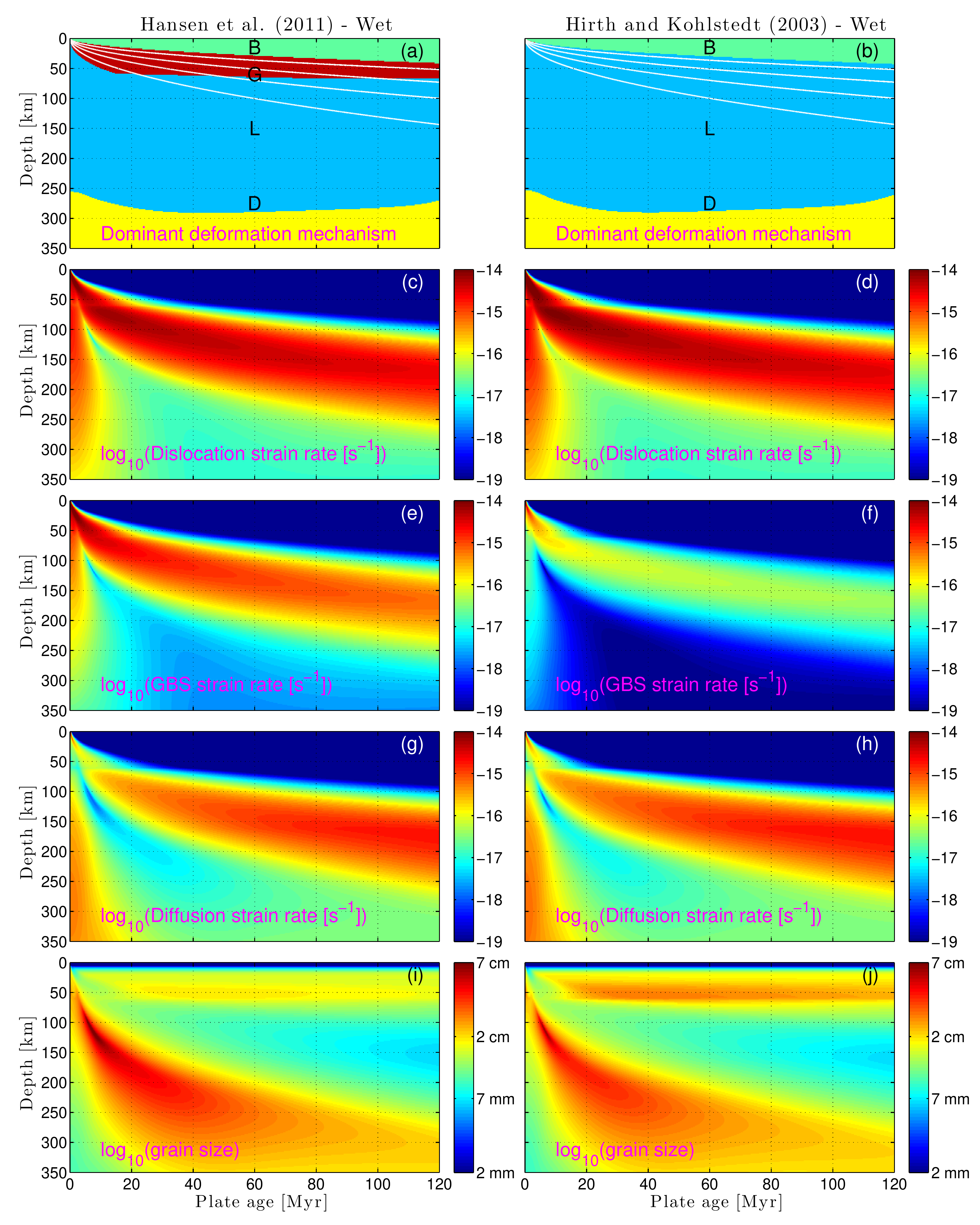}
  \caption{All panels are for a spreading-rate of $U_0 = 2$~cm/a. The
    left or right column use the \cite{Hansen11} or \cite{hirth03}
    parameters for grain boundary sliding. Panels (a) \& (b): The
    dominant deformation mechanism. Plasticity (B), diffusion creep
    (D), grain boundary sliding (G), and dislocation creep (L) are
    shown in green, yellow, red, and blue respectively. Isotherms are
    shown for $T = \left\lbrace 1200, 1000, 800, 600
    \right\rbrace^\circ$C in white. Panels (c) \& (d): $\log_{10}($
    Dislocation strain rate $[s^{-1}])$. Panels (e) \& (f):
    $\log_{10}($ GBS strain rate $[s^{-1}])$ . Panels (g) \& (h):
    $\log_{10}($ Diffusion strain rate $[s^{-1}])$. Panels (i) \& (j):
    $\log_{10}($ grain size $)$. }
  \label{fig:domin_wet}
\end{figure*}

Figure \ref{fig:domin_wet} shows the effect of our water
parametrisation upon the rheology and mean grain size field. The
values of $z_d$, $z_w$, and $C_{\text{OH}}^{\text{max}} $ are set as
57~km, 160~km and 1500 \waterunits{} respectively. As before, the two
columns of Figure~\ref{fig:domin_wet} differ only by the grain
boundary sliding parameters used; the left column uses \cite{Hansen11}
and the right column uses \cite{hirth03}.  The modification of
viscosity due to hydration results in accommodation of the
plate-driven deformation at greater depth (i.e., below $z_d =57 km$);
shallower than $z_d$, the total strain rate decreases. The effective
strain rate at depths less than $z_d$ are reduced by an order of
magnitude for the hydrous model compared to the dry model (compare
Figs.~\ref{fig:domin_HKH_map} and \ref{fig:domin_wet}), leading to an
increase in mean grain size at $z < z_d$ for the hydrous model.
Furthermore, because the plate-driven deformation is accommodated at a
greater depth, the region of large grain-size (III) approaches the
ridge axis more closely. This could be significant for the transport
of volatile rich partial melts.

\subsection{Sensitivity to parameters}
\label{sec:robust_params}

The preceding discussion explored grain-size dynamics with composite
rheology and hydration. We next focus on the predictions these models
make for global mid-ocean ridge spreading systems.  Specifically, the
ridge system is characterized by systematic variations in certain
parameters (e.g., spreading rate and potential temperature
\citep{gale13, dalton14}) and others that are simply uncertain due to
a lack of experimental or observational constraint (e.g., the
grain-growth exponent $p$ and mantle water content).

In this section we explore the sensitivity of the model to a subset of
these parameters. We consider the sensitivity of the mean grain-size
field to potential temperature, grain growth exponent, spreading rate,
and water concentration. Except where stated otherwise, parameters
used in the model are the same as for the reference case. The results
are presented in terms of a probability density function of the
asthenospheric grain size. The probability densities are constructed
from a region 100~km in depth and 200~km in width, centered on the
ridge axis, excluding places where the temperature is colder than
$600^\circ$C. This ensures that only regions of active creep are
included but still provides approximately 68,000 grid points over
which each probability density is calculated, for a grid spacing of
0.5~km.

\begin{figure*}
  \center
  \includegraphics[width = 12cm]{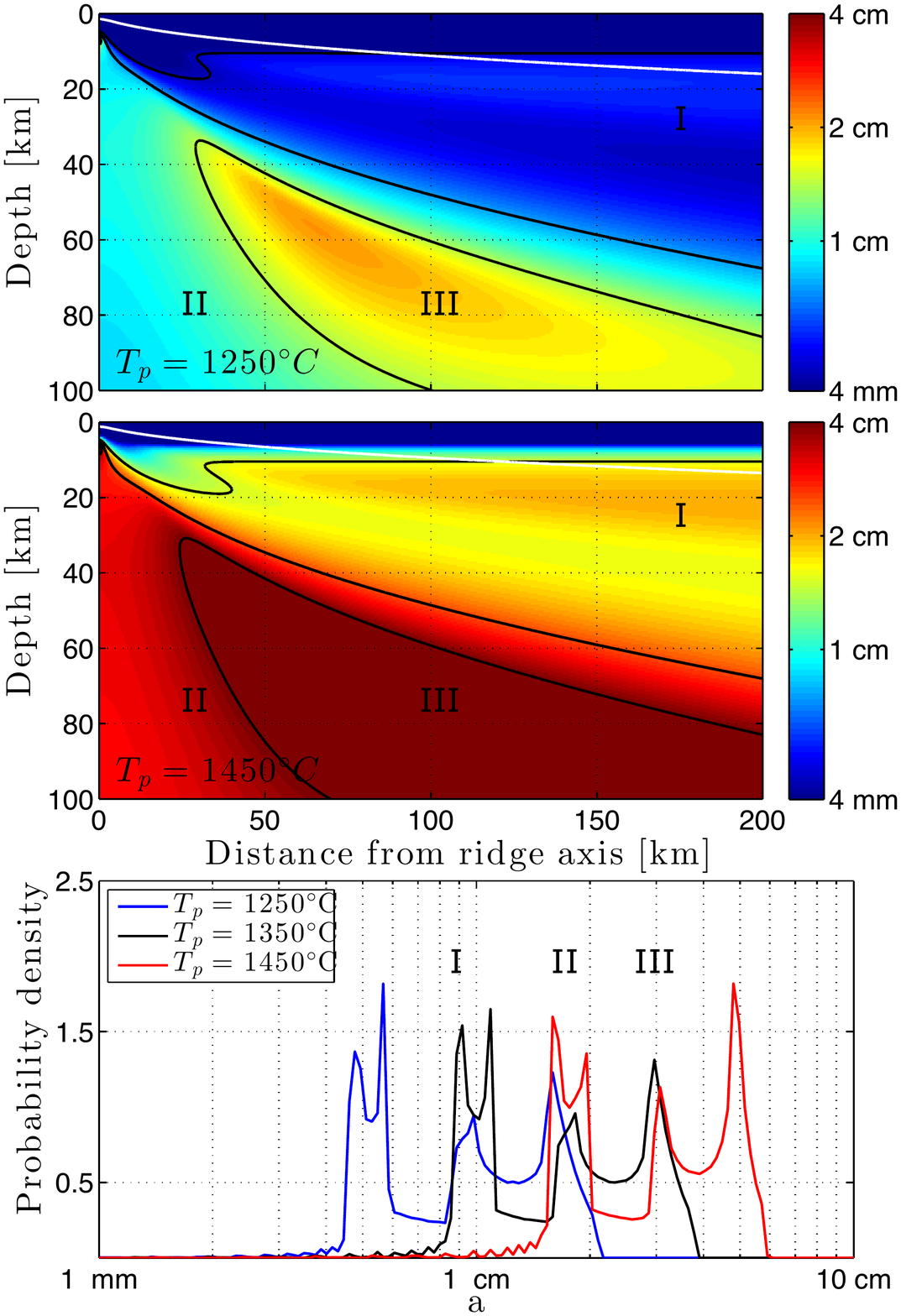}
  \caption{Sensitivity of mean grain size to potential temperature,
    $T_p$. Top panel: Mean grain size structure for $T_p =
    1250^\circ$C. Middle panel: Mean grain size structure for $T_p =
    1450^\circ$C. Bottom panel: Probability density for potential
    temperature $T = \{ 1250, 1350, 1450 \}^\circ$C in blue, black, and
    red respectively. Regions denoted by Roman numerals in the top and
    middle panel correspond to peaks in the probability density
    (bottom panel). }
  \label{fig:sens_potent}
\end{figure*}

We first examine the sensitivity of the mean grain-size field to
variations in the potential temperature
(Figure~\ref{fig:sens_potent}). The mean grain-size for the two
extreme cases (1250$^\circ$C and 1450$^\circ$C) are presented in the
top and middle panels with a white contour marking the 600$^\circ$C
isotherm. The probability density for each case is shown in the bottom
panel. The black curve corresponds to the reference potential
temperature (corresponding to the reference case shown in
Figure~\ref{fig:rep_fields}a) and potential temperatures of
1250$^\circ$C and 1450$^\circ$C are shown in blue and red,
respectively. The Roman numerals between black contours in the top and
middle panels correspond to equivalently labelled peaks of the
probability density functions.  However, the numerical values of grain
size along these contours are not equal between the top and middle
panels; rather, they correspond to equivalent spatial structures and
relate to peaks of probability density for the reference case (black
profile in bottom panel).

Variations in potential temperature approximately preserve the form of
the grain-size probability density functions. The change in grain
growth and reduction rates come from the Arrhenius dependence of
grain-boundary mobility and dislocation creep, respectively. The
present model uses an activation energy for grain boundary mobility of
350~kJ and for dislocation creep of 520~kJ; therefore the derivative
of grain growth rate with respect to temperature is always greater
than that of grain reduction. This difference results in a translation
of the grain-size probability density to smaller or larger grain sizes
for lower or higher potential temperatures, respectively.

\begin{figure*}
  \center
  \includegraphics[width = 12cm]{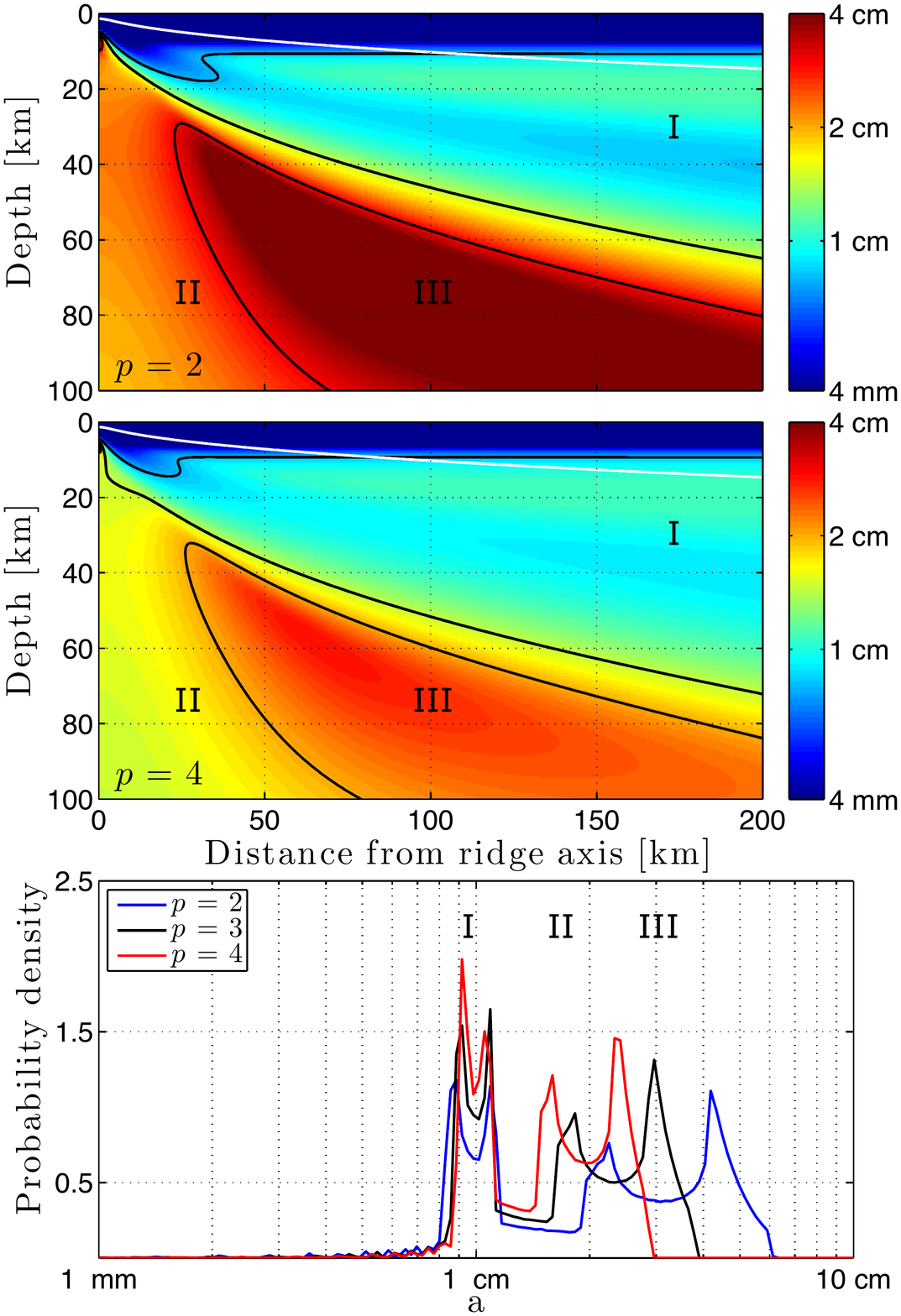}
  \caption{Sensitivity of mean grain size to grain growth exponent,
    $p$. Top panel: Mean grain size structure for $p = 2$. Middle
    panel: Mean grain size structure for $p = 4$. Bottom panel:
    Probability density for grain growth exponent $p = \{ 2, 3, 4 \}$
    in blue, black, and red respectively. Regions denoted by Roman
    numerals in the top and middle panel correspond to peaks in the
    probability density (bottom panel). }
  \label{fig:sens_expon}
\end{figure*}

We next investigate the sensitivity of the mean grain size field to
variations in the grain growth exponent $p$
(Figure~\ref{fig:sens_expon}). The mean grain size as a function of
space is plotted for grain growth exponent $p = \{2, 4 \}$ in the top
and middle panels, respectively.  Again the white contour is the
600$^\circ$C isotherm. The probability densities for each value of $p$
are shown in the bottom panel of Figure~\ref{fig:sens_expon}, where
the reference grain growth exponent is shown in black and
$p= \{2, 4 \}$ are shown in blue and red, respectively.  As before,
Roman numerals associate structures in the spatial domain with peaks
in the probability density functions.

The grain growth exponent determines the rate at which the mean grain
size grows, according to the relationship $a \propto t^{1/p}$. A
larger grain growth exponent therefore leads to a decreased growth
rate.  The coupling between both grain reduction and growth rates
through the mean grain size leads to an increase in mean grain size
only for those regions where grain growth rate dominates.  This is
seen clearly in the effect of $p$ on the probability density functions
in Figure~\ref{fig:sens_expon}.  Specifically, in Region I where grain
size reduction dominates, the peak of the distribution remains
relatively fixed at $a\approx1$~cm.  By contrast, in Regions II and
III, where grain growth is enhanced, smaller values of $p$ lead to
progressively larger grain sizes.  This contrasts with the case of
varying mantle potential temperature (Figure~\ref{fig:sens_potent}),
in which the Arrhenius term for reduction and growth rates are altered
in a complimentary fashion, and thus result in a translation (rather
than a stretching) of the probability densities.  Note that the
prefactor for grain growth is co-variant upon the other parameter
values, analogous to the viscous prefactor.  We make the assumption
that the rate of grain growth is unchanged for a reference grain size
$a_0$.  Therefore, $K_g$ is rescaled as follows,
\begin{equation*}
  K_g = K_{g\text{ref}} \frac{p}{p_\text{ref}}a_0^{p - p_\text{ref}}
\end{equation*}
where $K_{g\text{ref}}$ is the reference grain growth prefactor and
$p_\text{ref}$ is the reference grain growth exponent.

\begin{figure*}
  \center
  \includegraphics[width = 12cm]{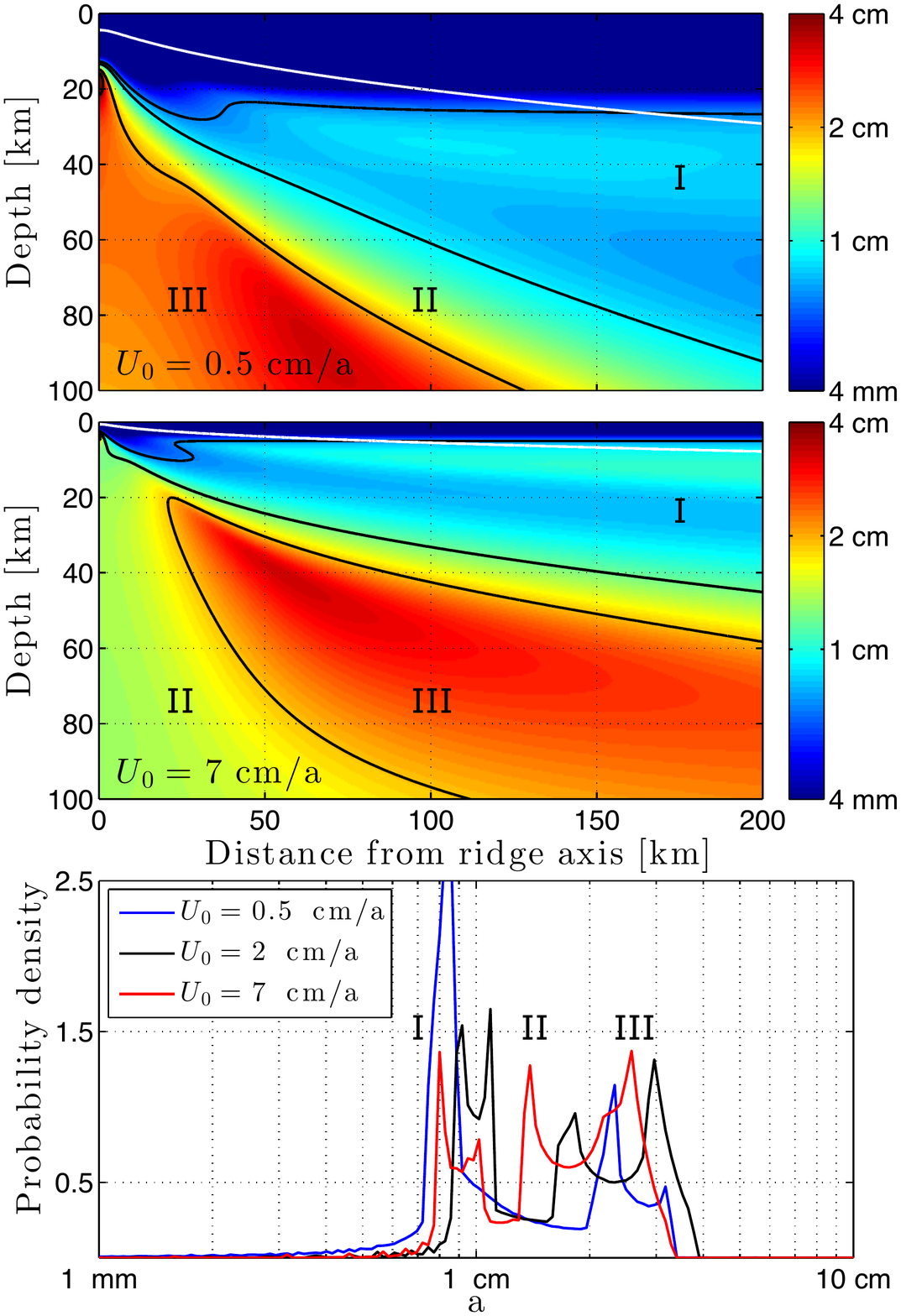}
  \caption{Sensitivity of mean grain size to spreading-rate, $U_0$.
    Top panel: Mean grain size structure for $U_0 = 0.5$~cm/a. Middle
    panel: Mean grain size structure for $U_0 = 7$~cm/a. Bottom panel:
    Probability density for spreading-rate
    $U_0 = \{ 0.5, 2, 7 \}$~cm/a in blue, black, and red
    respectively. Regions denoted by Roman numerals in the top and
    middle panel correspond to peaks in the probability density
    (bottom panel). }
  \label{fig:sens_spread}
\end{figure*}

We next evaluate the sensitivity of the mean grain-size field to
variations in spreading rate $U_0$ (Figure~\ref{fig:sens_spread}). The
mean grain size for $U_0 = \{ 0.5, 7 \}$~cm/a is shown in the top and
middle panels, respectively, and their probability density functions
are compared to the reference case with $U_0 = 2$~cm/a in the lower
panel.  The effect of spreading rate on the probability density can be
predicted from the nondimensional parameters $\mathcal{D}$ and
$\mathcal{G}$ (eqn.~(\ref{eq:grain_coefs})). These parameters suggest
that as the spreading rate is increased, the probability density will
undergo a translation to smaller grain size. This behaviour is evident
in a comparison of the probability densities of the slow and fast
spreading ridges. To understand the slowest spreading rate, we recall
that $U_0$ controls two fundamental properties of a mid-ocean ridge
system: the thermal profile of the adjacent lithosphere and the
magnitude of asthenospheric strain rates. A prediction based on
$\mathcal{D}$ and $\mathcal{G}$ assumes that the variation of strain
rate is of leading-order importance. This assumption is invalid for an
ultra-slow spreading rate; in that case, changes associated with the
thermal structure have a greater control on the distribution of mean
grain sizes.

\begin{figure*}
  \center
  \includegraphics[width = 12cm]{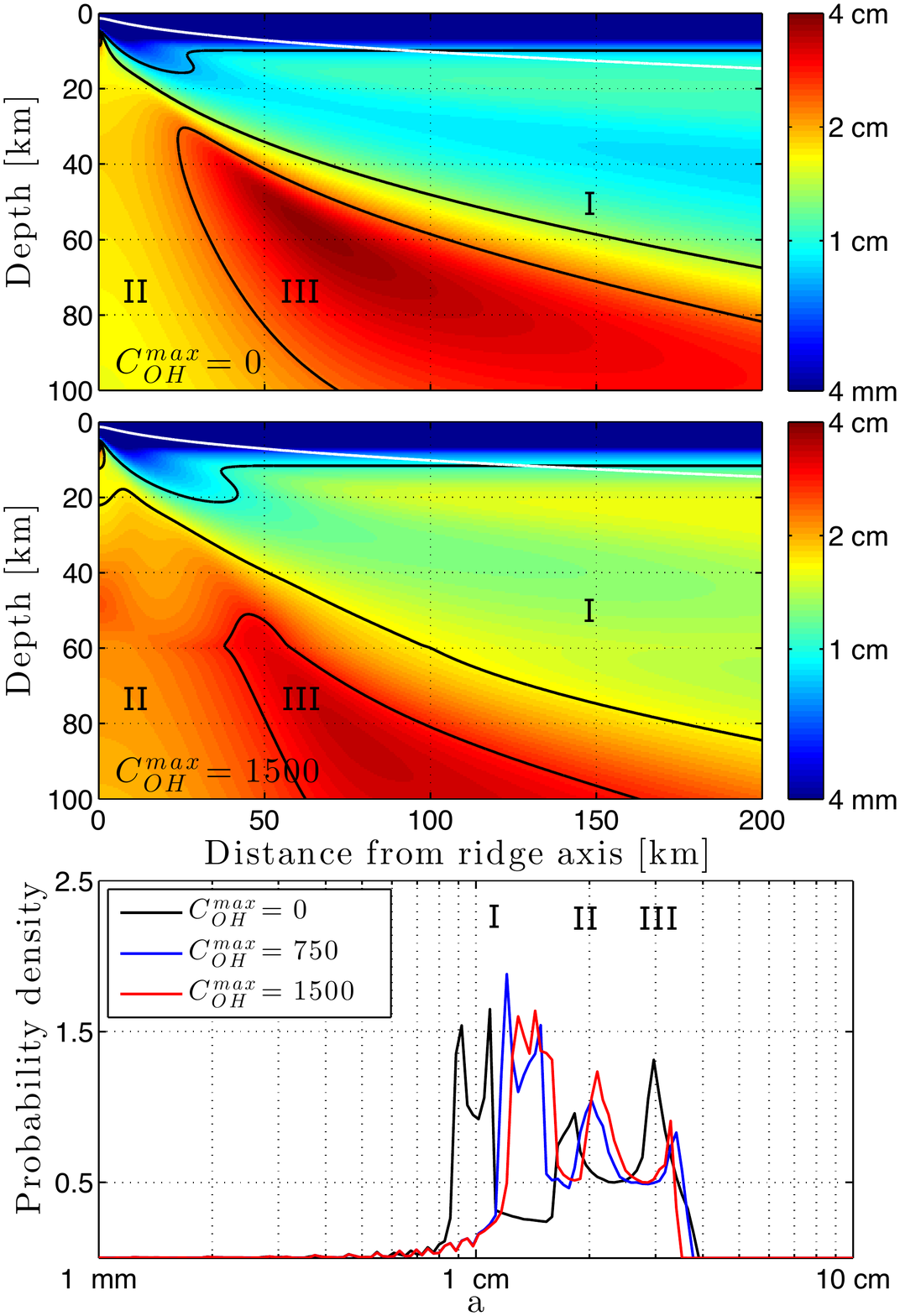}
  \caption{Sensitivity of mean grain size to water concentration,
    $C_{\text{OH}}^{\text{max}}$. Top panel: Mean grain size structure
    for $C_{\text{OH}}^{\text{max}} =0$ \waterunits{}. Middle panel:
    Mean grain size structure for $C_{\text{OH}}^{\text{max}} =1500$
    \waterunits{}. Bottom panel: Probability density for spreading-rate
    $C_{\text{OH}}^{\text{max}} =\{0, 750, 1500 \}$ \waterunits{} in
    black, blue, and red respectively. Regions denoted by Roman
    numerals in the top and middle panel correspond to peaks in the
    probability density (bottom panel). }
  \label{fig:sens_coh}
\end{figure*}

Lastly, we investigate the sensitivity of the mean grain-size field to
variations in the deep-asthenospheric water concentration
$C_{\text{OH}}^{\text{max}}$ (Figure~\ref{fig:sens_coh}). The spatial
distribution of mean grain size for
$C_{\text{OH}}^{\text{max}}= \{0, 1500 \}$ \waterunits{} is shown in
the top and middle panels, respectively. The probability density
funcitons are shown in the bottom panel, where
$C_{\text{OH}}^{\text{max}} = \{0, 750, 1500 \}$ \waterunits{} are
shown in black, blue, and red, respectively.

In our model, water content enters the grain-size evolution
implicitly, by reduction of viscosity; therefore, only the grain-size
reduction rate is altered by water. This is analogous to the
sensitivity of grain size to $p$, in that only a single rate prefactor
in equation (\ref{eq:grain_rate_nondi}) is explicitly altered. Hence,
one expects to see a contraction/dilation of the probability density
at the small grain-size side of the distribution, which is indeed
evident in the bottom panel of Figure~\ref{fig:sens_coh}.

The mean grain-size probability densities for
$C_{\text{OH}}^{\text{max}} =750$ and $1500$ \waterunits{} are very
similar. This is to be expected, given the dominance of dislocation
creep in the present models. Under dislocation creep, the stress
exponent and water exponent are $n_L = 3.5$ and $r_L = 1.2$,
respectively. The dependence of dislocation creep rate on water
concentration thus scales as $C_{\text{OH}}^{r_L/n_L}$. Therefore, the
change in viscosity due to water content is greatest for the first few
hundred $C_{\text{OH}}$; viscosity is weakened to approximately 80\%
of its original value when the water concentration is increased from
$C_{\text{OH}}= 100$ to $200$ or $C_{\text{OH}}= 750$ to $1500$
\waterunits{}.

\section{Permeability of the partially molten region}
\label{sec:perm}

A key objective of this paper is to predict the influence of mean
grain size on the permeability structure of the mantle beneath a
mid-ocean ridge.  Mantle permeability at low melt fraction may be
written as a function of mean grain size $a$ and melt fraction $\phi$
as $K = a^2 \phi^n/c$, where $c$ and $n$ are empirically determined
constants related to the geometry of the pore network
\citep[e.g.][]{mckenzie84, vonbargen86}. Recent work by
\cite{miller14} indicates that $n\approx2.6$ and $c\approx60$ are
appropriate for mantle conditions.  This value for $c$ is
significantly lower than previous estimates \citep{wark98}, except for
that of \cite{Connolly09}, who found much higher permeability overall
and suggested $c\sim3$--$30$.  For consistency with previous work on
two-phase magma dynamics, we choose $c=500$ and $n=3$.  It is
straightforward to rescale these results for smaller $c$; permeability
becomes larger overall.

The current model considers only a single-phase mantle, and hence does
not offer an obvious means for computing melt fraction that is
consistent with conservation of mass, momentum, and energy. Therefore,
in order to estimate the dimensional permeability, we use the solidus
parameterisation from \cite{katz03} (eqn.~(4)) to determine the region
in which temperature and pressure conditions are favourable for the
stability of partial melt. Within this region, we make the simplifying
assumption of a constant melt fraction $\phi=0.01$ to emphasize the
contribution of grain size on the permeability structure. Outside of
the region of partial melting, permeability is set to zero.

\begin{figure*}
  \center
  \includegraphics[width = 12cm]{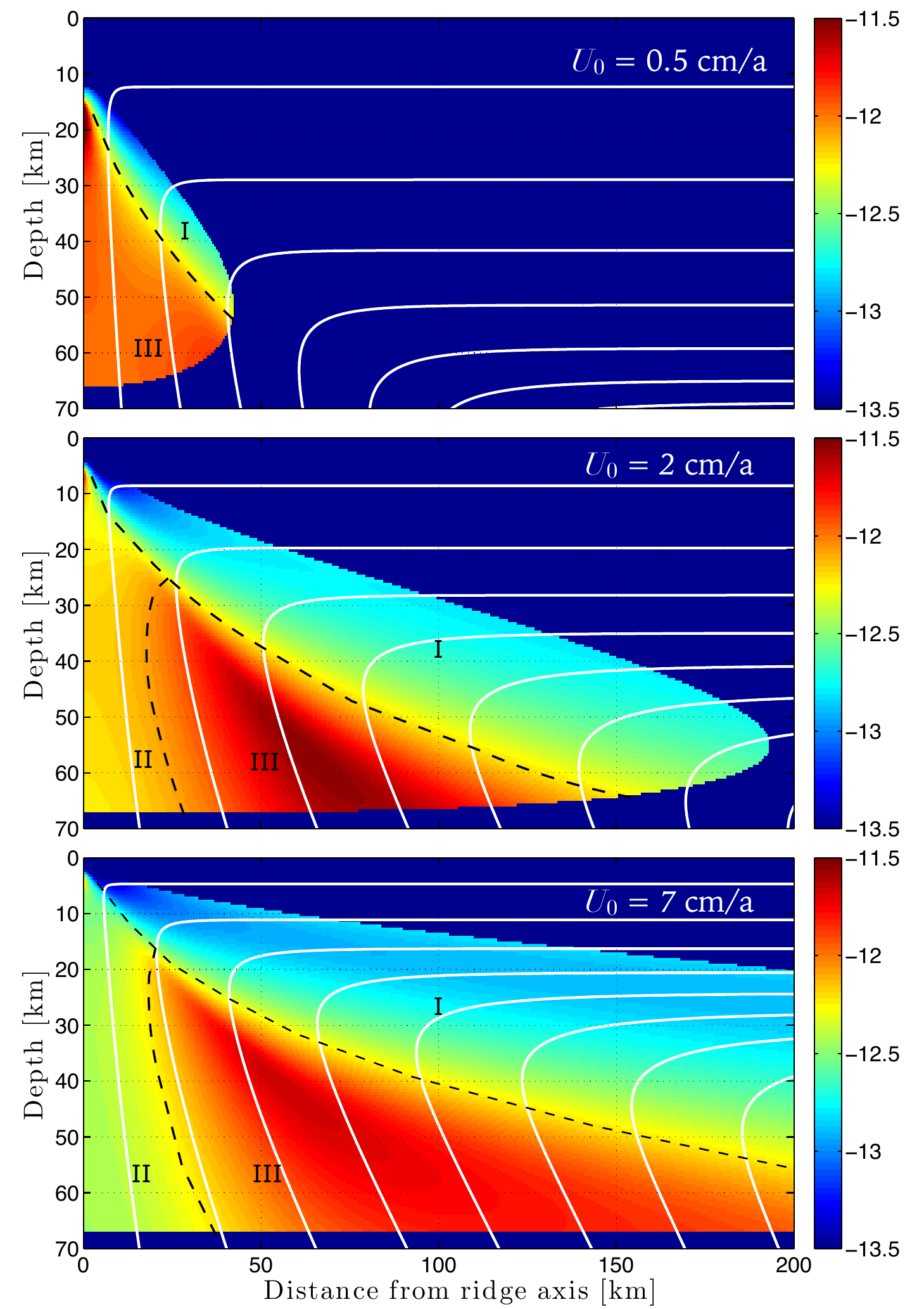}
  \caption{Dimensional permeability, $\log_{10}( K [\text{m}^2])$,
    assuming a constant porosity of 1\%. The colour scale for
    permeability has been truncated to
    $\log_{10}( K [\text{m}^2]) \in [-11.5, -13.5]$. A spreading-rate
    of $U_0 = \{0.5, 2, 7 \}$~cm/a is shown in the top, middle, and
    bottom panel respectively.  Regions of the permeability structure
    (I, II, III) are separated by dashed black lines (see
    text). Mantle streamlines are shown in white. }
  \label{fig:perm_vel_comp}
\end{figure*}

The predicted dimensional permeability for spreading rates of
$U_0 = \{ 0.5, 2, 7 \}$~cm/a are shown in the top, middle, and bottom
panels of Figure~\ref{fig:perm_vel_comp}, respectively. The colour-bar
is truncated to show a dimensional permeability of
$\log_{10}(K [\text{m}^2]) \in [-11.5, -13.5]$.  Mantle streamlines
are shown in white.  For purposes of discussion, each panel has been
split into three regions: region I has a low permeability compared to
the rest of the melting regime; region II has intermediate
permeability and is approximately columnar beneath the ridge axis; and
region III has the largest permeability of the melt region. These
regions are approximately the same as were used to describe grain size
in section~\ref{sec:robust_params}.

The middle panel of Figure~\ref{fig:perm_vel_comp} displays the
permeability structure for the reference model. This may be broadly
described as a higher permeability mantle (regions II \& III) beneath
lower permeability mantle (region I). The transition from high to low
permeability follows a curve that slopes upward toward the ridge
axis. The grain-size induced change of permeability across this
transition is about one order of magnitude.

The top and bottom panels of Figure~\ref{fig:perm_vel_comp} show
simulations for an ultra-slow ($U_0=0.5$~cm/a) and fast ($U_0=7$~cm/a)
spreading rate, respectively. The primary effect of varying the
spreading rate is to alter the thermal structure within the domain,
which in turn alters the extent of the partially molten region. A
secondary effect is to alter the magnitude of the maximum strain rate
beneath the ridge, shifting the balance between grain growth and grain
size reduction. The net result of thermal and strain-rate effects is
to increase the significance of region I with increasing spreading
rate.  At faster spreading rate, the permeability of region I
decreases and occupies a larger fraction of the partially molten
region.

Vertical, extensional strain rates in the column of mantle beneath the
ridge axis control the grain size and permeability of Region II. At
slow spreading, extensional strain rates in this region are minimal,
allowing grains to grow to large size and permeability to
increase. Under these conditions, shown in the top panel of
Fig.~\ref{fig:perm_vel_comp}, region II is not discernable and
essentially merges with region III. With increasing spreading rate,
however, larger extensional strain rates lead to differentiation of
region II from region III by reduction of grain size.  The distinction
between regions II and III is greatest at fast spreading rate (bottom
panel Fig.~\ref{fig:perm_vel_comp}).

Region III is evident at all spreading rates shown in
Figure~\ref{fig:perm_vel_comp}. This region is characterized by very
slow deformation rates and high temperatures. The mean grain size is
stabilized at its maximum values in this location, yielding the
highest permeability there.

Although melt transport is not included in the present model, it is
interesting to speculate on how modification of the permeability
structure by variations in grain size might affect melt migration.
Calculations shown in \cite{katz08b} provide a reference case for melt
flow in a mid-ocean ridge setting with grain size assumed to be
constant (see Fig.~4a of \cite{katz08b}).  As predicted by
\cite{sparks91}, melt tends to rise vertically under buoyancy until it
reaches the permeability barrier associated with sub-solidus
temperatures in the overlying lithosphere.  At the permeability
barrier, melt is deflected toward the ridge axis and travels through a
high-permeability channel located immediately below the barrier.  The
key feature of the \cite{sparks91} model is that there exists an
upward transition from non-zero to zero permeability along a barrier
that is sloping with respect to the horizontal. \cite{spiegelman93c}
considered the efficiency of lateral deflection of melt in this
context, relating it to the sharpness of the freezing front and the
compaction length in the region below it.

The gradient in permeability at the bottom of region I in
Figure~\ref{fig:perm_vel_comp} is not as sharp as that associated with
the freezing front at the base of the lithosphere, but it may
nonetheless function in an analogous way.  At the base of region I,
permeability is high and the compaction length (at $\phi=0.01$) is
large, due to higher mean grain-size there.  Above the transition,
permeability is reduced (though note that there is no difference in
potential temperature across the transition).  Melt can penetrate
across this gradient, but increasing Darcy drag and consequent
compaction pressure associated with the lower permeability might
deflect melt laterally toward the ridge axis, leading to preferential
migration along the base of region I.  Therefore, the grain-size
induced transition of permeability may be thought of as a ``soft''
permeability barrier, giving rise to a variant of \cite{sparks91}-type
melt focusing.

Given the structure of mantle flow, melt production is more rapid
below region I where mantle upwelling, and thus melt production rates,
are greater.  If the soft barrier at the base of region I is
effective, melts could accumulate there, increasing porosity above the
constant value of 1\% assumed here.  This would in turn increase
permeability, potentially resulting in a sloping decompaction channel
\citep{sparks91}. The presence of this channel would sharpen the
permeability contast at base of region I, and its steep slope would
resolve a large component of the buoyancy force along it (larger than
that along the more shallowly sloping barrier at the bottom of the
lithosphere).

If a ``soft'' permeability barrier associated with grain-size
variations is capable of efficiently focusing magma toward the ridge
axis, we would expect low porosity throughout region I of the model.
In contrast, the mantle in the triangular zone above the dry solidus
and below region I would host more melting and higher porosity.  It
might therefore stand out in seismic or magnetotelluric inversions as
a steep-sided region of slow shear-wave speed or higher electrical
conductivity.  It is interesting to compare this hypothesis with
recent magnetotelluric (MT) observations.  MT studies by \cite{Baba06}
and \cite{Key13} along the East Pacific Rise imaged triangular regions
of high conductivity with sides sloping downward at about 45$^\circ$
to the (horizontal) spreading direction.  The observed slope is
signficantly greater than that calculated based on the porosity field
computed in two-phase flow models \citep{katz08b}, which predict a
much shallower slope coinciding with the base of the oceanic thermal
boundary layer.  However, these MT observations could be explained by
a soft permeability barrier associated with grain-size variations in
the melting region.  Specifically, a barrier formed at the base of
region I for the fast-spreading case shown in
Figure~\ref{fig:perm_vel_comp} would produce a steep sided, highly
conductive triangle similar to that observed in the MT data.  Future
calculations that fully couple grain size evolution and two-phase flow
are needed to explicitly test these predictions and determine whether
the ``soft'' barrier would generate an efficient mechanism to
channelise rising melts toward the ridge axis.

\section{Conclusion}
\label{sec:conc}

In this study we have presented a two-dimensional, single phase model
for the steady-state mean grain size beneath a mid-ocean ridge. The
model employs a composite rheology incorporating diffusion creep,
dislocation creep, grain boundary sliding, and plasticity.  Mean grain
sizes were calculated using the paleowattmeter model of
\cite{AustEv07}.

We investigated the robustness of the mean grain-size field to
variations in the grain boundary sliding parameters by comparing the
experimentally determined parameters of \cite{Hansen11} and
\cite{hirth03}. It was found that the structure of the mean grain-size
field is generally insensitive to grain boundary sliding parameters.

We also investigated the robustness of the mean grain-size field to
mantle hydration state.  We imposed a one-dimensional parametrisation
of mantle water concentration and coupled this concentration into the
dynamics through the viscosity terms only. Interestingly, the presence
of water had the greatest impact on the mean grain-size at depths less
than approximately 60~km.  This was due to a shift in the location of
maximum strain rate to greater depth as a consequence of the more
compliant, hydrated mantle below the dehydration boundary.

We considered the sensitivity of the mean grain-size field to
variations of parameters in the grain evolution model.  If a parameter
influences both the growth and reduction-rate prefactors, then the
mean grain-size probability density may undergo a translation to
larger or smaller grain sizes. This was observed for the case of
potential temperature, and when comparing the slow spreading rate with
a fast spreading rate. In contrast, for parameter variations that only
influence either the growth or reduction rate prefactor, we obtain a
stretching of the mean grain-size probability density at the large or
small grain-size side of the distribution. This was seen in the case
of the grain growth exponent and water concentration.

Finally, we studied the impact of the mean grain-size field on the
permeability structure for a half spreading-rate of
$U_0 = \{ 0.5, 2, 7 \}$~cm/a. We assumed a uniform melt fraction of
1\% within the expected melt region and found that, for all spreading
rates, the permeability structure due to mean grain size may be
approximated as a high permeability region beneath a low permeability
region. The transition between high and low permeability regions forms
a boundary that is steeply sloped toward the ridge axis. This is, to
some extent, analogous to the permability barrier often hypothesized
to form at the base of the lithosphere and we suggest that it may
similarly focus melt towards the ridge axis.  This focusing may, in
turn, constrain the region where significant melt fractions are
observed by seismic or magnetotelluric surveys. This interpretation of
melt focusing via the grain-size permeability structure is consistent
with MT observation of the asthenosphere beneath the East Pacific Rise
\citep{Baba06, Key13}.

We emphasize that these predictions for permeability and melt focusing
beneath mid-ocean ridges are based on results from a single-phase
model formulation for solid mantle flow. The incorporation of a liquid
magmatic phase is expected to alter the results.  Due to the more
complex coupling between viscosity, flow, melting, porosity,
grain-size, and permeability under two-phase flow, it is difficult to
predict how such a model would differ from the results and predictions
obtained here. A more detailed investigation of mid-ocean-ridge
grain-size dynamics, in the context of two-phase, coupled magma/mantle
dynamics, is the focus of forthcoming work.

\section{Acknowledgements}

The authors acknowledge helpful discussions with Brian Evans, Lars
Hansen, Kerry Key, Tobias Keller, Sander Rhebergen, and John Rudge.  
The research leading to these results has received funding from the European
Research Council under the European Union's Seventh Framework
Programme (FP7/2007-2013) / ERC grant agreement number 279925. Katz is
grateful for additional support from the Leverhulme Trust.
Parameter values used to produce the numerical results are
provided in Tables \ref{tab:viscsymbols}, \ref{tab:grainsymbols}, and in the text.


\begin{table*}[ht]
  \centering
  \begin{tabular}{llll}
    Symbol & Units & Description \\
    \hline
    $\vel$ & cm a$^{-1}$ & velocity  \\
    $P$ & Pa & Dynamic Pressure  \\
    $\bar{P}$ & Pa & Total Pressure  \\
    $\eta$ & Pa s & Viscosity  \\
    $\dot{ \boldsymbol{ \varepsilon}}$ & s$^{-1}$ & strain-rate tensor \\
    $\dot{  \varepsilon}$ & s$^{-1}$ & second invariant of strain-rate \\
    $ \boldsymbol{ \sigma}$ & Pa & deviatoric stress tensor \\
    $ \sigma$ & Pa & second invariant of deviatoric stress \\
    $\dot{ \boldsymbol{ \sigma}}$ & Pa s$^{-1}$ & stress-rate tensor \\
    $T$ & K & Temperature \\
    $x_r$ & km & Ridge width \\
    $k$ & - & deformation mechanism index \\
    $\sigma_Y$ & Pa & Yield stress \\
    $t$ & s & time \\
    $\dot{W}$ & J s$^{-1}$ & work-rate \\
    $V$ & m$^3$ & Volume \\
    $S$ & m$^2$ & Surface area \\
    $\mathcal{E}$ & J m$^{-3}$ & Energy per unit volume \\
    $U_0$ & cm a$^{-1}$ & Spreading-rate \\
    $T_p$ & $^\circ$C & Potential temperature \\
    $C_{\text{OH}}$ & \waterunits{} & Water concentration \\
    \hline
  \end{tabular}
  \caption{Mathematical notation and units.}
  \label{tab:variables}
\end{table*}

\bibliographystyle{plainnat}
\bibliography{manuscript}
\end{document}